\documentclass[a4paper,fleqn,usenatbib]{mnras}

\usepackage{mathptmx}

\usepackage[T1]{fontenc}
\usepackage{ae,aecompl}

\usepackage{graphicx}	
\usepackage{amsmath}	
\usepackage{amssymb}	

\newcommand{\cmnt}[1]{}
\newcommand{\review}[1]{#1}

\newcommand{\software}[1]{\texttt{\sc #1}}

\newcommand{\starname}{TIC\,378898110}

\newcommand{\porb}{$P_\mathrm{orb}$}

\newcommand{\mdot}{$\dot{M}$}

\newcommand{\hei}{\ion{He}{I}}
\newcommand{\heii}{\ion{He}{II}}

\newcommand{\filus}{\textit{u$_s$}}	
\newcommand{\filgs}{\textit{g$_s$}}	
	
\newcommand{\filis}{\textit{i$_s$}}	
	
\newcommand{\filu}{\textit{u'}}	
\newcommand{\filg}{\textit{g'}}	
	
\newcommand{\fili}{\textit{i'}}

\newcommand{\mpia}{$^1$}
\newcommand{\tlv}{$^2$}
\newcommand{\warwick}{$^3$}
\newcommand{\boston}{$^4$}
\newcommand{\highpoint}{$^5$}
\newcommand{\unc}{$^6$}
\newcommand{\riogrande}{$^7$}
\newcommand{\valparaiso}{$^8$}
\newcommand{\virginia}{$^{9}$}
\newcommand{\morehead}{$^10$}
\newcommand{\cork}{$^{11}$}
\newcommand{\manchester}{$^{12}$}
\newcommand{\ing}{$^{13}$}
\newcommand{\ardastella}{$^{14}$}
\newcommand{\uw}{$^{15}$}
\newcommand{\ioa}{$^{16}$}
\newcommand{\sheffield}{$^{17}$}
\newcommand{\iac}{$^{18}$}
\newcommand{\michigan}{$^{19}$}

\title[A Bright AM\,CVn Binary in \textit{TESS}]{TIC 378898110: A Bright, Short-Period AM\,CVn Binary in \textit{TESS}}

\author[M. J. Green et al.]{
Matthew J. Green\mpia$^,$\tlv$^,$\warwick\thanks{E-mail: mjgreenastro@gmail.com},
J. J. Hermes\boston,
Brad N. Barlow\highpoint$^,$\unc,
T. R. Marsh\warwick,
\newauthor
Ingrid Pelisoli\warwick,
Boris T. G\"{a}nsicke\warwick,
Ben C. Kaiser\unc,
Alejandra Romero\riogrande,
\newauthor
Larissa Antunes Amaral\riogrande$^,$\valparaiso,
Kyle Corcoran\virginia,
Dirk Grupe\morehead,
Mark R. Kennedy\cork$^,$\manchester,
\newauthor
S. O. Kepler\riogrande,
James Munday\warwick$^,$\ing,
R. P. Ashley\warwick,
Andrzej S. Baran\ardastella$^,$\uw,
\newauthor
Elm\'{e} Breedt\ioa,
Alex J. Brown\sheffield,
V. S. Dhillon\sheffield$^,$\iac,
Martin J. Dyer\sheffield,
\newauthor
Paul Kerry\sheffield,
George W. King\michigan$^,$\warwick,
S. P. Littlefair\sheffield,
Steven G. Parsons\sheffield,
\newauthor
and David I. Sahman\sheffield.
\\
\mpia \review{Max-Planck-Institut f\"{u}r Astronomie, K\"{o}nigstuhl 17, D-69117 Heidelberg, Germany}
\\
\tlv School of Physics and Astronomy, Tel-Aviv University, Tel-Aviv 6997801, Israel
\\
\warwick Astronomy and Astrophysics Group, Department of Physics, University of Warwick, Coventry, CV4 7AL, United Kingdom
\\
\boston Department of Astronomy, Boston University, 725 Commonwealth Ave., Boston, MA 02215, USA
\\
\highpoint Department of Physics and Astronomy, High Point University, High Point, NC, USA
\\
\unc Department of Physics and Astronomy, University of North Carolina, Chapel Hill, NC, USA
\\
\riogrande Instituto de F\'{i}sica, Universidade Federal do Rio Grande do Sul, Av. Bento Goncalves 9500, Porto Alegre 91501-970, RS, Brazil
\\
\valparaiso Instituto de F\'{i}sica y Astronom\'{i}a, Universidad de Valpara\'{i}so, Gran Breta\~{n}a 1111, Playa Ancha, Valpara\'{i}so 2360102, Chile
\\
\virginia University of Virginia, Department of Astronomy, 530 McCormick Rd., Charlottesville, VA 22904, USA
\\
\morehead Department of Physics, Geology, and Engineering Technology, Northern Kentucky University, Highland Heights, KY 41076, USA
\\
\cork School of Physics, University College Cork, Cork, Ireland
\\
\manchester Jodrell Bank Centre for Astrophysics, Department of Physics and Astronomy, The University of Manchester, Manchester M13 9PL, UK
\\
\ing Isaac Newton Group of Telescopes, Apartado de Correos 368, E-38700 Santa Cruz de La Palma, Spain
\\
\ardastella ARDASTELLA Research Collaboration, Missouri State University, Springfield, MO\,65897, USA
\\
\uw Astronomical Observatory, University of Warsaw, Al. Ujazdowskie 4, 00-478 Warszawa, Poland
\\
\ioa Institute of Astronomy, University of Cambridge, Madingley Road, Cambridge, CB3~0HA, United Kingdom
\\
\sheffield Department of Physics and Astronomy, University of Sheffield, Sheffield, S3 7RH, United Kingdom
\\
\iac Instituto de Astrof\'isica de Canarias, 38205 La Laguna, Tenerife, Spain
\\
\michigan Department of Astronomy, University of Michigan, Ann Arbor, MI 48109, USA
}

\date{Accepted XXX. Received YYY; in original form ZZZ}

\pubyear{2023}

\begin{document}
\label{firstpage}
\pagerange{\pageref{firstpage}--\pageref{lastpage}}
\maketitle

\begin{abstract}
AM\,CVn-type systems are ultracompact, helium-accreting binary systems which are evolutionarily linked to the progenitors of thermonuclear supernovae and are expected to be strong Galactic sources of gravitational waves detectable to upcoming space-based interferometers.
AM\,CVn binaries with orbital periods $\lesssim 20$--23\,min exist in a constant high state with a permanently ionised accretion disc.
We present the discovery of TIC\,378898110, a bright ($G=14.3$\,mag), nearby ($309.3 \pm 1.8$\,pc), high-state AM\,CVn binary discovered in \textit{TESS} two-minute-cadence photometry.
At optical wavelengths this is the third-brightest AM\,CVn binary known. 
The photometry of the system shows a $23.07172(6)$\,min periodicity, which is likely to be the `superhump' period and implies an orbital period in the range 22--23\,min.
There is no detectable spectroscopic variability.
The system underwent an unusual, year-long brightening event during which the dominant photometric period changed to a shorter period (constrained to $20.5 \pm 2.0$\,min), which we suggest may be evidence for the onset of disc-edge eclipses.
The estimated mass transfer rate, $\log (\dot{M} / \mathrm{M_\odot} \mathrm{yr}^{-1}) = -6.8 \pm 1.0$, is unusually high and may suggest a high-mass or thermally inflated donor.
The binary is detected as an X-ray source, with a flux of \review{$9.2 ^{+4.2}_{-1.8} \times 10^{-13}$\,erg\,cm$^{-2}$\,s$^{-1}$} in the 0.3--10\,keV range.
TIC\,378898110 is the shortest-period binary system discovered with \textit{TESS}, and its large predicted gravitational-wave amplitude makes it a compelling verification binary for future space-based gravitational wave detectors.
\end{abstract}



\begin{table}
\caption{Summary of basic observational properties of \starname.
The UVOT magnitude is given in the AB system, while others are given in the Vega magnitude system (as is customary). Sources for these magnitudes are given in Section~\ref{sec:sed}.}
\begin{center}
\begin{tabular}{lr}
\label{tab:basic}
Property & Value \\
\hline
TIC ID & 378898110\\
\textit{Gaia} ID & 6058834949182961536\\
ICRS coords.\ (J2000) & 12:03:38.7 $-$60:22:48.0 \\
Galactic coords. & 297.055664 +1.945349 \\
\\
\textit{Magnitudes:}\\
UVOT $UVW2$ & $14.41 \pm 0.03$\\
\textit{Gaia} $BP$ & $14.271 \pm 0.006$\\
\textit{Gaia} $G$ & $14.276 \pm 0.003$\\
\textit{Gaia} $RP$ & $14.243 \pm 0.006$\\
\textit{2MASS} $J$ & $14.11 \pm 0.04$\\
\textit{2MASS} $H$ & $14.08 \pm 0.05$\\
\textit{2MASS} $K_s$ & $14.11 \pm 0.07$\\
\textit{WISE} $W1$ & $14.20 \pm 0.06$\\
\textit{WISE} $W2$ & $14.40 \pm 0.08$\\
\hline
\end{tabular}
\end{center}
\end{table}

\section{Introduction}

AM\,CVn-type binary systems are ultracompact, mass-transferring binary systems with orbital periods in the range 5--68\,minutes \citep[e.g.][]{SolheimAMCVn,Green2020}. 
Each AM\,CVn binary consists of a white dwarf accreting helium-dominated matter from a degenerate or semi-degenerate donor star.
Short-period AM\,CVn binary systems are expected to be \review{among the first handful of} individual Galactic sources of gravitational waves in the frequency range visible to space-based interferometers such as the Laser Interferometer Space Antenna \citep[\textit{LISA};][]{Kremer2017,Breivik2018,Kupfer2018,Kupfer2023}.
Their evolutionary channels are linked to double white dwarf binaries, a significant channel for Type Ia Supernovae \citep[e.g.][]{Bildsten2007,Kilic2014,Maoz2014}.

A variety of progenitor channels have been proposed for AM\,CVn binaries, in which the system may descend from a double white dwarf binary \citep{Paczynski1967,Deloye2007,Wong2021b}, a binary consisting of a white dwarf and a compact helium-burning star \citep{Savonije1986,Iben1987,Yungelson2008}, or a cataclysmic variable with an evolved donor \citep{Podsiadlowski2003,Goliasch2015,Belloni2023}.
The relative importance of these various channels remains an unsolved problem.

AM\,CVn binaries remain a rare class of object, with 56 known systems at the time of the last population review \citep{Ramsay2018}.
Recent discoveries by \citet{Burdge2020}, \citet{vanRoestel2021a,vanRoestel2022}, \citet{RiveraSandoval2021a}, and others, have increased the number of published systems to approximately 80, but many newly discovered systems are not suitable for in-depth characterization studies (for instance, using phase-resolved spectroscopy) due to their faintness.

The small sample size of known AM\,CVn binaries limits attempts to understand the population empirically, such as studies of the population density \citep{Carter2013} or investigations into the question of how these systems form \citep{Goliasch2015,Green2018b,Ramsay2018,Wong2021b,Belloni2023}.
Based on the \textit{Gaia} parallaxes of the known systems, the sample of known AM\,CVn binaries may be incomplete even at distances of a few hundred parsecs \citep{Ramsay2018}.


AM\,CVn accretion discs show a range of types of photometric behaviour, commonly explained by the disc instability model that also describes hydrogen-accreting cataclysmic variables \citep[e.g.][]{Tsugawa1997,Cannizzo2015}.
This behaviour is driven by the mass transfer rate, \mdot, which correlates steeply with orbital period, \porb, as \mdot\,$\propto$\,\porb$^{-5.2}$ \citep[][]{Warner1995,Tsugawa1997}.
If the population includes donor stars with differing levels of degeneracy \citep{Deloye2007,Wong2021b} then some amount of scatter can be expected in the relationship between \porb\ and \mdot, but the correlation should remain strong.

As a result, AM\,CVn binaries can be separated into several groups based on their accretion disc behaviour, which naturally sorts them as a function of orbital period \citep[e.g.][]{SolheimAMCVn}.
Short-period systems (7--12 $\lesssim$\,\porb\,$\lesssim$ 20--28\,min) typically exist in a permanently ionised `high' state, comparable to nova-like cataclysmic variables. 
Meanwhile, systems at somewhat longer periods (20--28 $\lesssim$\,\porb\,$\lesssim$ 45--58\,min) spend the majority of their time in a neutral, quiescent state, with occasional dwarf nova outbursts. 
While most AM\,CVn binaries with \porb\,$> 20$\,min are outbursting or quiescent systems, two high-state systems have claimed orbital periods of 23\,min and 28\,min \citep[CXOGBS\,J1751-2940 and ZTF\,J2228+4949, though note that neither orbital period has been confirmed spectroscopically;][]{Wevers2016,Burdge2020b}.

High state AM\,CVn binaries and AM\,CVn binaries in outburst can show a variety of photometric signatures.
Assuming there are no eclipses, the strongest signal is typically a `superhump' signal, which is driven by an interaction between the donor star and the accretion disc, and has a period within a few per cent of the orbital period \citep{Patterson1993}.
The orbital period itself and the precession period of the accretion disc may be seen \citep[e.g.][]{Armstrong2012,Green2018b,Solanki2021}, and some systems also show a variety of periodic signals with no conclusive physical interpretation \citep[e.g.][]{Fontaine2011,Kupfer2015}.

In this work we present the discovery of the high-state AM\,CVn binary TIC\,378898110, which has properties presented in Table~\ref{tab:basic}.
The system was first identified as a short-period variable by its \textit{TESS} photometry \citep[Transiting Exoplanet Survey Satellite,][]{Ricker2014}.
Follow-up observations from the ground led to its AM\,CVn classification.
Its magnitude of $G=14.3$ makes it the third-brightest AM\,CVn binary known by apparent magnitude, after HP\,Lib ($G=13.6$) and the namesake of the class, AM\,CVn itself ($G=14.0$).
Its parallax from \textit{Gaia} EDR3 \citep{GaiaCollaboration2021} implies a distance of $306.2 \pm 1.7$\,pc \citep{Bailer-Jones2021}, making it the 12th closest AM\,CVn binary. 
The likely orbital period of TIC\,378898110 is $\approx 22$--23\,min, placing it in the overlap period range between high-state and outbursting systems.

In Section~\ref{sec:observations} we describe the observations undertaken for this work. 
Section~\ref{sec:photometry} presents the photometric data obtained and its analysis, while Section~\ref{sec:spectroscopy} presents the spectroscopic data.
Section~\ref{sec:discussion} and Section~\ref{sec:conclusions} discuss and summarise our findings regarding \starname.

\begin{table*}
\caption{Summary of the observations presented in this work. For ULTRACAM observations, exposure times in brackets are those used for the \filus-band, which are increased to account for the lower throughput of that filter. Approximate wavelength coverages of photometric bandpasses are given, to aid the reader in comparison between different bands.}
\begin{tabular}{lcccc}
\label{tab:observations}
Instrument & Date & Filters / Wavelength [\AA] & Exposure [s] & Total exposure [s]\\
\hline
\textit{Space-based phot.}\\
\textit{TESS} & 2021 Apr 02--May 26 & \textit{T} / 6000--10\,000 & 120 & 4138200\\
\textit{Swift} & 2021 July 27 & \textit{UVW2} / 1600--3500 & -- & 1240\\
\\
\textit{Ground-based phot.}\\
Goodman+SOAR & 2021 July 4 & \textit{S8612} / 3000--8400 & 10 & 7660\\
ULTRACAM+NTT & 2021 July 14 & \filus\filgs\filis\ / 3000--8700 & 2.8 (8.4) & 6300\\
ULTRACAM+NTT & 2022 March 05 & \filus\filgs\filis\ / 3000--8700  & 3.5 (10.5) & 3570\\
ULTRACAM+NTT & 2022 March 06 & \filus\filgs\filis\ / 3000--8700  & 3.0 (9.0) & 4080\\
ULTRACAM+NTT & 2022 March 28 & \filus\filgs\filis\ / 3000--8700  & 3.0 (9.0) & 4580\\
ULTRACAM+NTT & 2023 March 08 & \filus\filgs\filis\ / 3000--8700  & 3.0 (6.0) & 9790\\
ULTRACAM+NTT & 2023 March 09 & \filus\filgs\filis\ / 3000--8700  & 3.0 (6.0) & 2230\\
ULTRACAM+NTT & 2023 March 10 & \filus\filgs\filis\ / 3000--8700  & 3.0 (6.0) & 4020\\
\\
\textit{Spectroscopy}\\
Goodman+SOAR & 2021 July 05 & 3600--5200 & 60 & 2940\\
\hline
\end{tabular}
\end{table*}

\section{Observations}
\label{sec:observations}

A number of observations of \starname\ were obtained and analysed for this work. 
These observations are summarised in Table~\ref{tab:observations}.

\subsection{\textit{TESS} Photometry}

\starname\ was observed by \textit{TESS} with a two-minute cadence in Sectors 37 and 38, spanning a total coverage of 52\,days. 
The target was proposed for two-minute cadence data under proposals G03124 (as a candidate low-mass white dwarf) and G03221 \citep[as a candidate hot subdwarf, due to its inclusion in the hot subdwarf catalogue of][]{Geier2019}.
These data were reduced by the Science Processing Operations Centre (SPOC).

The target is located in a crowded region of the sky, with an estimated \textit{TESS} contamination factor (\texttt{CROWDSAP}) of 0.12; the relative amplitudes from SPOC lightcurves have been corrected to account for this flux dilution. 
A pixel-level analysis \citep{Higgins2023} showed that the periodic variability was likely to originate from \starname\ itself, as was later confirmed by ground-based photometry.

A previous \textit{TESS} observation in Sectors 10 and 11 (2019 March to May) had only a cadence of half an hour and was not able to fully resolve the dominant periodic signal.

\subsection{Ground-Based Photometry}


Follow-up photometry was obtained with the Goodman High Throughput Spectrograph \citep{Clemens2004} on the 4.1\,m Southern Astrophysical Research (SOAR) telescope at Cerro Pach\'{o}n in Chile.
These data were obtained using a broad, blue \textit{S8612} filter. 
We obtained 443 exposures of 10\,s, taken in the 200~Hz \texttt{ATTN2} readout mode with $2 \times 2$ binning and a reduced window on the chip to minimize readout overheads. 
The data were debiassed and flat-field corrected with standard \software{iraf} routines, and aperture photometry was performed with \software{daophot}.

Further ground-based photometry was obtained using ULTRACAM, a high-speed, triple-beam photometer \citep{ULTRACAM}. 
For these observations, ULTRACAM was mounted on the 3.5\,m New Technology Telescope (NTT) at La Silla observatory in Chile. 
The \filus, \filgs, and \filis\ filters were used; these filters are designed to cover the same wavelengths as the Sloan \filu\filg\fili\ filters but with higher throughput \citep{HIPERCAM}.
The wavelength coverage of these filters can be broken down into 3000--4000\,\AA\ for \filus, 3900--5700\,\AA\ for \filgs, and 6700--8700\,\AA\ for \filis.
Observations were obtained in 2021, 2022 and 2023.

The ULTRACAM data were reduced using the HiPERCAM pipeline \citep{HIPERCAM}.
Each image was bias-subtracted and divided throughout by a flat-field image in the same filter that was obtained at twilight on the same night.
No dark-frame subtraction was performed, but care was taken to avoid known hot pixels during target acquisition.
The target flux was extracted using a variable aperture with width scaled to 1.7$\times$ the full-width half-maximum of the point-spread function in that image.
The target flux was divided by a constant comparison star observed in the same image (coordinates 12:03:35.19 $-$60:23:06.9, $G$=14.1\,mag) to correct for changes in atmospheric transparency.

\subsection{Swift X-Ray Observations}
\label{sec:swift}

Motivated by the presence of a \textit{ROSAT} source within 15\arcsec\ of the target (1RXS J120340.6-602252)\footnote{Notable given that the position uncertainty of \textit{ROSAT} sources can be as large as 16\arcsec\ for low-S/N sources \citep{Ayres2004}.}, X-ray observations were undertaken with the NASA Neil Gehrels Gamma-Ray Burst Exploer Mission
\textit{Swift} satellite \citep{gehrels2004} through its Target of Opportunity program (ToO ID 16045).
The target was observed by \textit{Swift} for 1240\,s.
During these observations the X-Ray Telescope \citep[XRT;][]{burrows2005} was in photon counting mode \citep{hill2004}. Source counts were selected in a circular region with a radius of 30\arcsec. Background counts were extracted from a source-free nearby region with a radius of 212\arcsec\ applying \software{xselect}. Using the exposure map, an auxiliary response file was created with the tool \software{xrtmkarf}. The spectral data were analyzed without rebinning in \software{xspec} \citep{arnaud1996} and using Cash statistics \citep{cash1979}.

The UltraViolet and Optical Telescope \citep[UVOT;][]{roming2005} was in event mode with the \textit{UVW2} filter (mode 0x0121). The data for source and background regions were extracted in circles with radii of 7\arcsec\ and 20\arcsec, respectively. Magnitudes and flux densities were determined using \software{uvotsource} with the calibration as described in \citet{poole2008} and \citet{breeveld2010}.

\subsection{Spectroscopy}

Phase-resolved spectroscopy with 1\,min exposures and a cadence of $\approx$65.5\,s was collected using the Goodman High Throughput Spectrograph on the SOAR telescope \citep{Clemens2004}.
A volume phase holographic (VPH) grating with 930\,lines per mm was used, giving a wavelength coverage of approximately 3700--5200\,\AA. 
The slit width was 1\,arcsec, giving a resolution of 2.9\,\AA\ (resolving power $\approx$1500).
These data covered approximately one hour, bracketed by Fe arc lamp exposures before the first spectrum, after half an hour, and after the final spectrum. The spectra were reduced using custom \software{python}--based tools and an optimal extraction routine based on the methods described by \citet{Marsh1989}.

A comparison star (Gaia EDR3 6058834949186192768, magnitude $G$=16.2) was also on the slit and observed simultaneously with the target.
The comparison star did not have sufficient S/N for analysis within individual exposures. 
A fit to five Balmer lines in the summed spectrum of the comparison star found that they were consistent with their rest wavelengths ($12 \pm 16$\,km s$^{-1}$).

\section{Photometric Analysis}
\label{sec:photometry}
\subsection{TESS}

\begin{figure*}
\includegraphics[width=2\columnwidth]{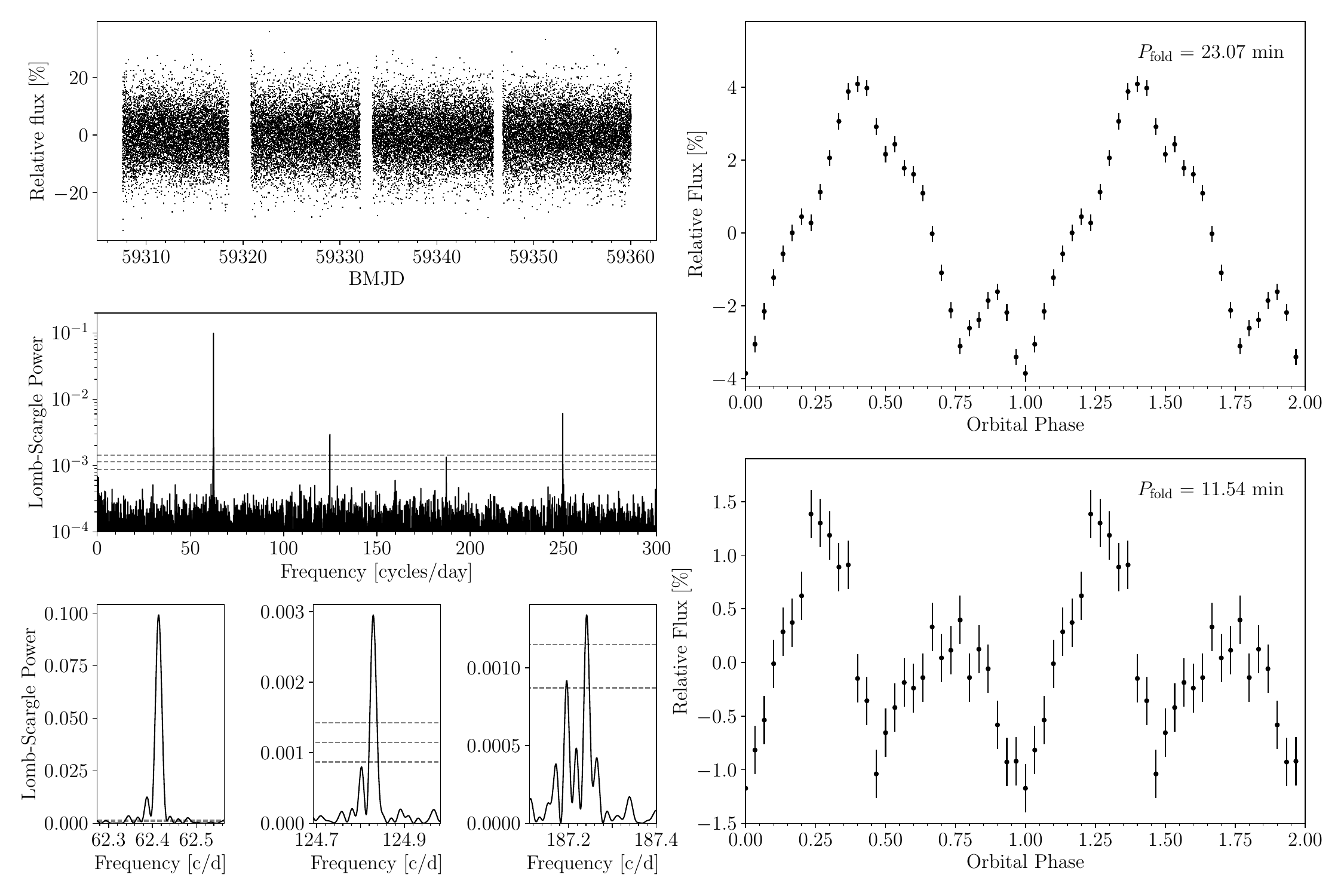}
\caption{
\textit{Top left:} \textit{TESS} lightcurve from Sectors 37 and 38, showing flux change relative to the mean.  
\textit{Middle left:} Lomb-Scargle periodogram of the \textit{TESS} lightcurve. Dashed horizontal lines show false alarm probabilities of 0.1, $10^{-3}$, and $10^{-5}$.
\textit{Lower left:} Zoomed plots of the periodogram peaks around the 23.07\,min period and its first two harmonics, with the same false alarm probabilities marked (c/d refers to cycles per day). Note the change in $y$-axis scale from logarithmic to linear.
\textit{Top right:} Phase-folded and binned \textit{TESS} lightcurve, folded on the 23.07\,min period.
\textit{Lower right:} Phase-folded and binned \textit{TESS} lightcurve, folded on the first harmonic at 11.5\,min.
}
\label{fig:tess}
\end{figure*}

\begin{figure}
\includegraphics[width=\columnwidth]{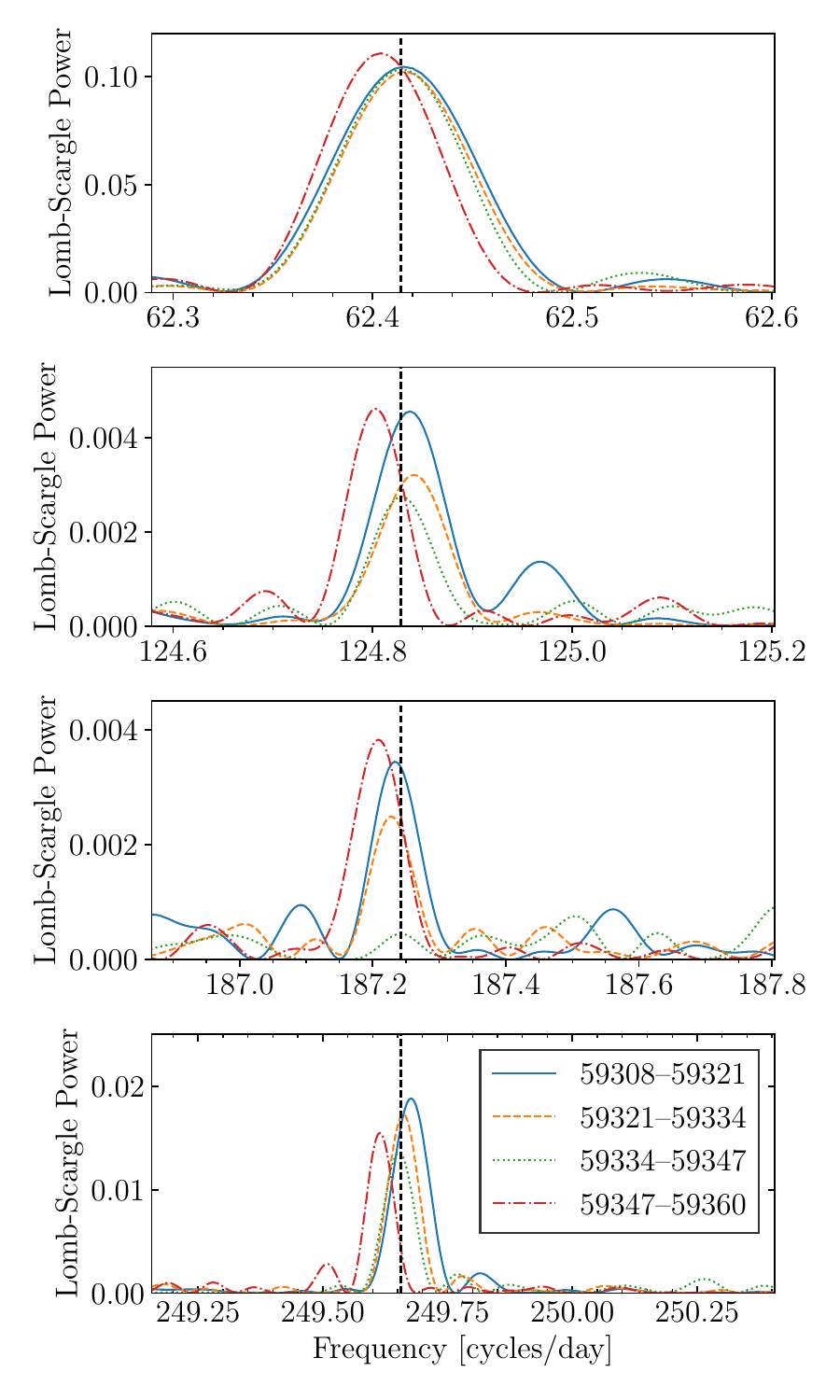}
\caption{Lomb-Scargle periodograms of \textit{TESS} data, after splitting the data into four equal segments of approximately 13 days each (BJD ranges are denoted in the figure legend). \review{Each row shows one of the four strongest harmonics of the 23.07\,min signal. Note that the axis range in each panel has been scaled by the harmonic number, so that a consistent offset in the figure represents a consistent fractional change in frequency between different harmonics.}
The vertical dashed line shows the 23.07\,min signalderived from the combined periodogram of the entire \textit{TESS} lightcurve \review{and its harmonics}.
The periodic signal appears to be constant through the first three segments, while it appears to drift towards a slightly lower frequency during the final segment.
}
\label{fig:tess-split-fold}
\end{figure}

In Fig.~\ref{fig:tess} we show the \textit{TESS} lightcurve of \starname. 
A Lomb-Scargle periodogram \citep{Lomb1976,Scargle1982} shows a strong peak with a period of 23.07172(6)\,min, as well as weaker peaks at harmonics of that signal.
The uncertainty on the period was characterised by fitting a sine wave to the data.

When phase-folded on the 23.07\,min period (Fig.~\ref{fig:tess}, top right panel), the \textit{TESS} data show non-sinusoidal, sawtooth-like variability, in which the rise is steeper than the decline, which is characteristic of superhump variability in high-state or outbursting AM\,CVn binaries \citep[e.g.][]{Armstrong2012,Green2018b}.
The bump feature just before minimum light resembles a feature seen in superhumps of CR Boo early during its outburst \citep[the `Stage A' superhumps; e.g.][]{Isogai2016}.
We also show a phase-fold of the second harmonic of this period, 11.5\,min, which gives a somewhat similar form with a weaker amplitude and lower significance.

In order to test for any change in the photometric period, we split the \textit{TESS} lightcurve into four equal-length segments of approximately 13 days each.
Fig.~\ref{fig:tess-split-fold} shows the Lomb-Scargle periodograms of \review{the first four harmonics} in each of these segments. 
The period appears constant through the first three segments. 
However, in the fourth segment, the period appears to drift towards a somewhat lower frequency. 
\review{The strengths of the higher harmonics change more significantly than the fundamental, but the variation in their frequencies is comparable.}
After characterising the uncertainties on the fundamental period in each segment by fitting a sine wave to the data, we found that the period drift was significant at a 3-$\sigma$ level compared to the previous segments.
The measured frequencies were 62.415(2), 62.417(2), 62.413(2), and 62.405(2) cycles day$^{-1}$, respectively, for the four segments.

In the periodogram in Fig.~\ref{fig:tess}, we note a marginal peak with a separation of $\approx 0.05$ per cent at the low-frequency wing of the 23.07\,min periodicity and each of its harmonics.
This is likely an imprint of the slightly lower frequency seen towards the end of the \textit{TESS} observations.


Unlike in some previously studied systems with space-based photometry \citep{Green2018b,Solanki2021}, we do not see a low-frequency signal resulting from the accretion disc precession in \starname.

\subsection{Ground-Based Photometry}

\begin{figure}
\includegraphics[width=\columnwidth]{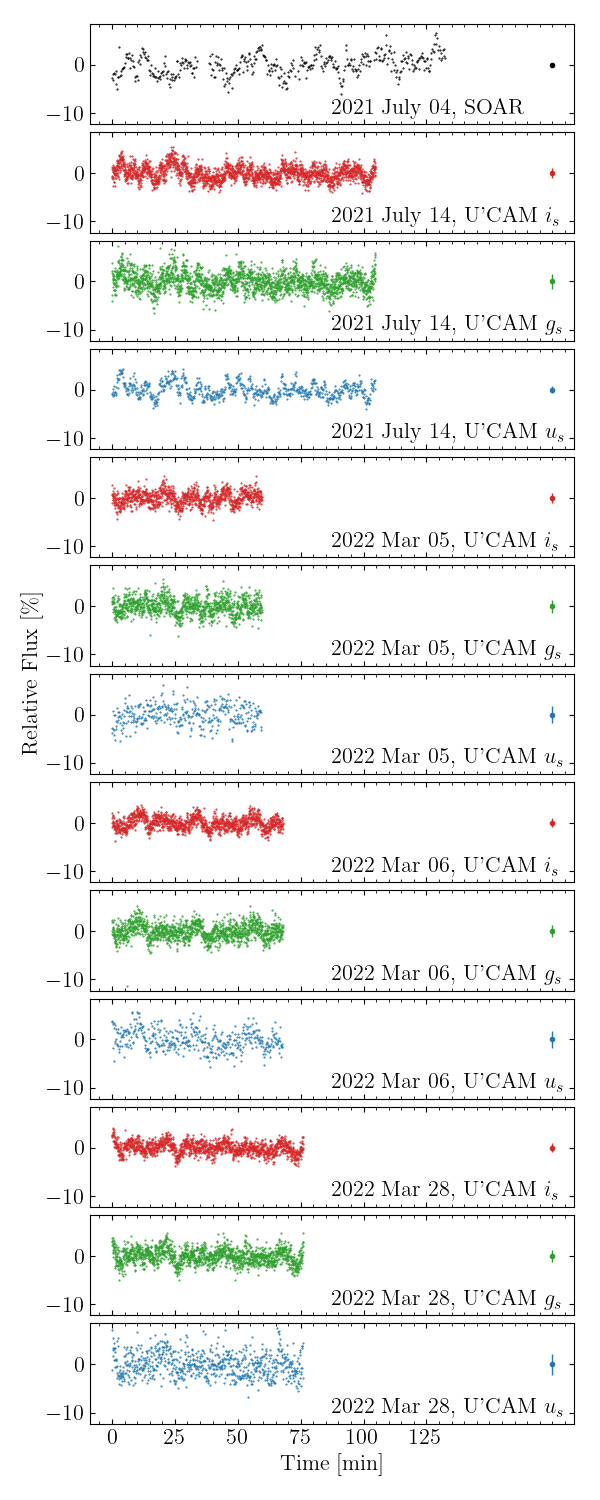}
\caption{Lightcurves of \starname\ obtained through an \textit{S8612} filter with Goodman+SOAR and through \filus\filgs\filis\ with ULTRACAM+NTT in 2021 and 2022.
Right-hand data point shows the typical uncertainty.
}
\label{fig:ground}
\end{figure}

\begin{figure}
\includegraphics[width=\columnwidth]{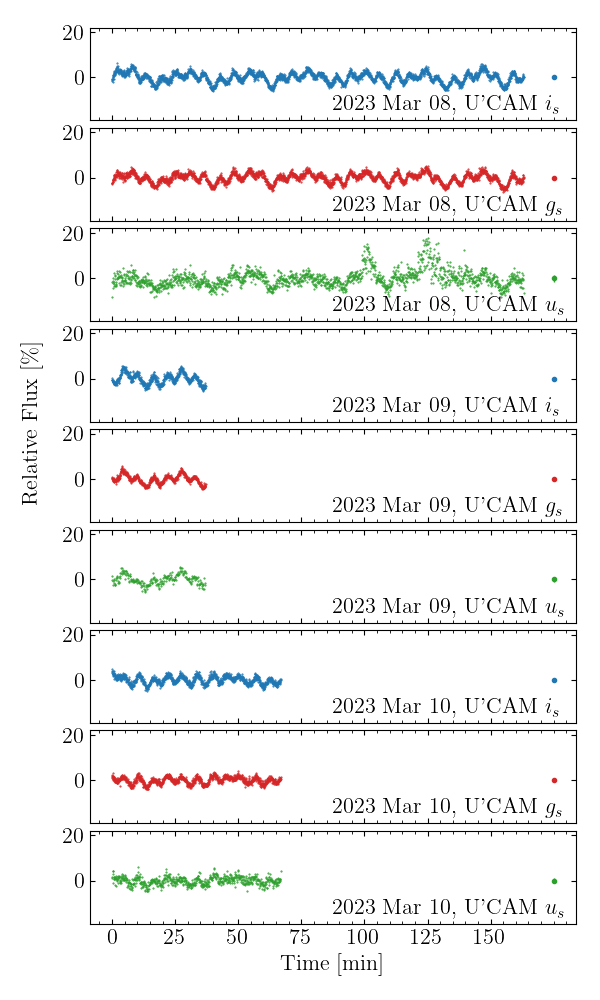}
\caption{Lightcurves of \starname\ obtained through \filus\filgs\filis\ filters with ULTRACAM+NTT in 2023.
}
\label{fig:ground-2}
\end{figure}

\begin{figure}
\includegraphics[width=\columnwidth]{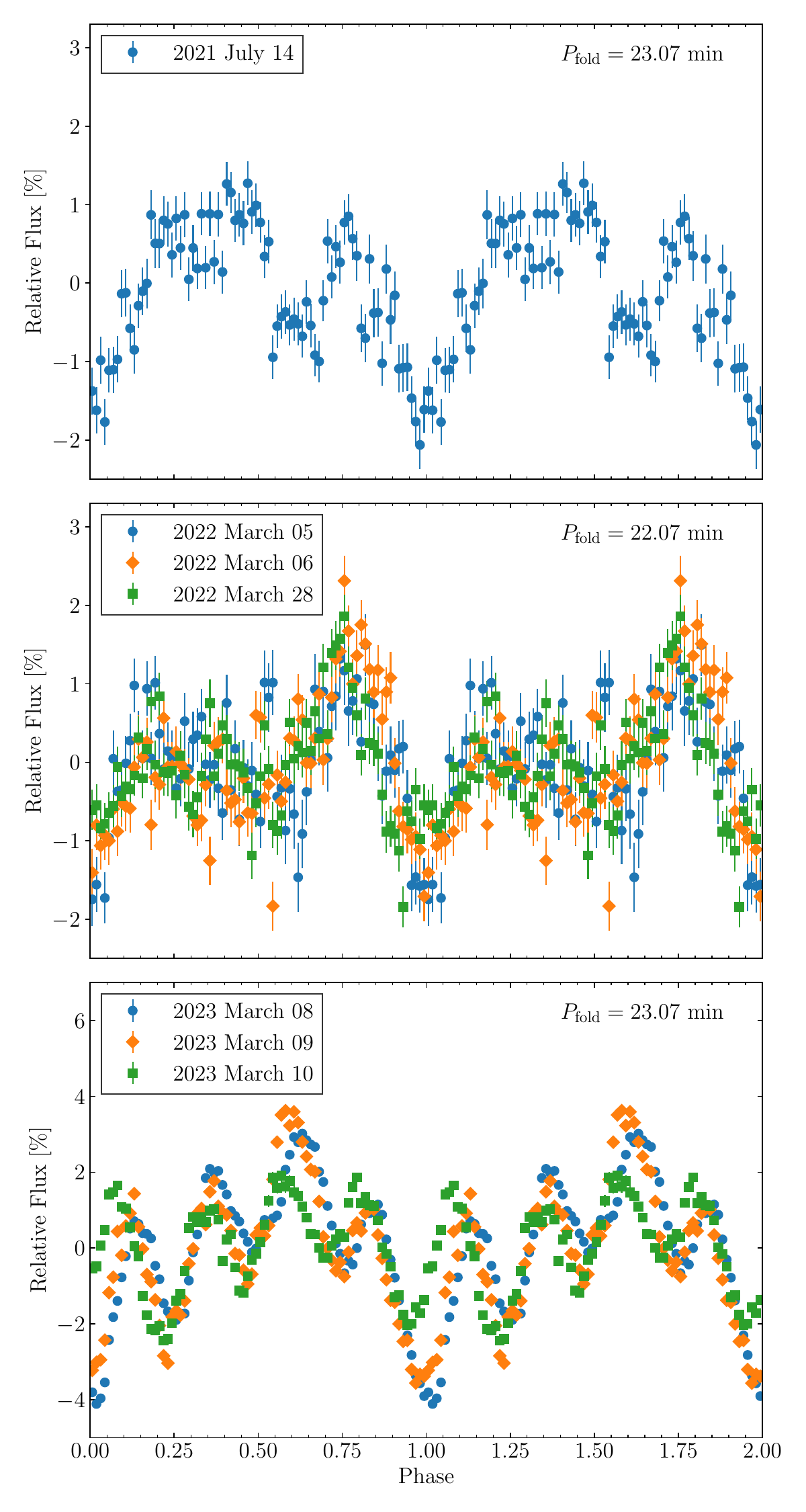}
\caption{
Phase-folded \filgs-band ULTRACAM lightcurve of \starname\ from 2021 (top), 2022 (middle), and 2023 (bottom).
Within each panel, data from multiple nights were folded with the same period and phase offset.
The zero-phase was chosen as the phase of minimum light.
In 2022, folding the data on the \textit{TESS} photometric period of 23.07\,min did not reconcile the phase of minimum light between the three nights, and so a different folding period of 22.07\,min was used.
Note the change in \textit{y}-axis scale between panels.
The significant improvement in SNR in 2023 was due to the larger amplitude of the signal and an improvement in atmospheric seeing for the 2023 nights.
}
\label{fig:ground-folded}
\end{figure}

\begin{figure}
\includegraphics[width=\columnwidth]{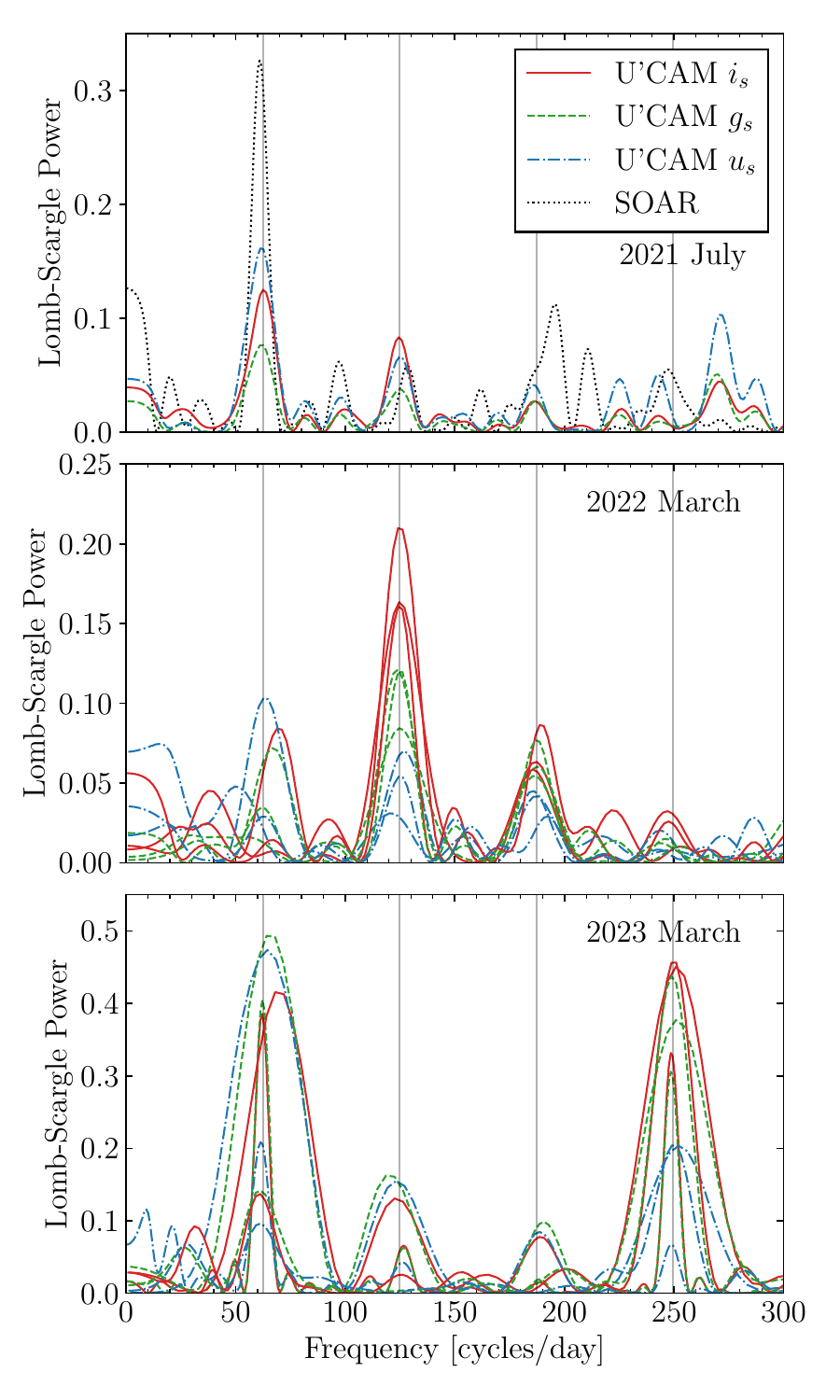}
\caption{
Lomb-Scargle periodograms of the lightcurves in Figs.~\ref{fig:ground}--\ref{fig:ground-2}, separated into data from 2021, 2022 and 2023. 
Vertical lines mark the first four harmonics of the \textit{TESS} frequency (23.07\,min).
The strength of the second harmonic in 2022 and the fourth harmonic in 2023 are significantly increased relative to the fundamental frequency.
}
\label{fig:ground-lomb}
\end{figure}

\begin{figure}
\includegraphics[width=\columnwidth]{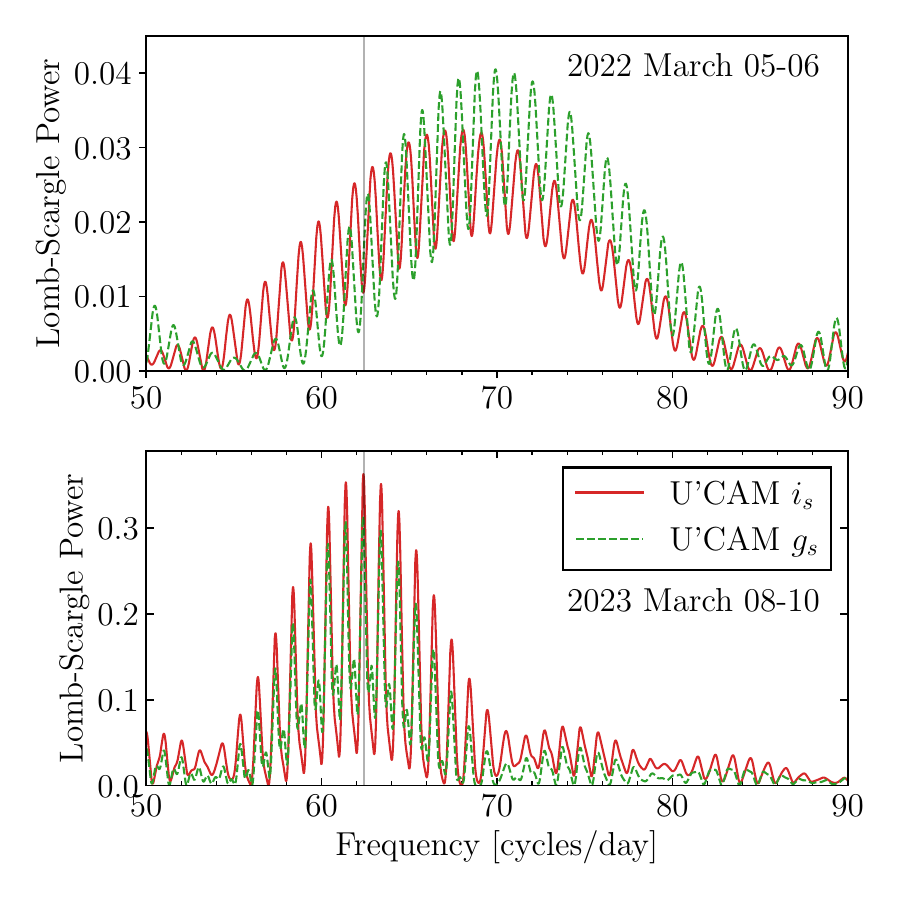}
\caption{Combined Lomb-Scargle periodograms of multiple nights from 2022 and 2023.
The vertical line marks the dominant \textit{TESS} frequency (23.07\,min).
}
\label{fig:ground-lomb-zoom}
\end{figure}

Figs.~\ref{fig:ground}, \ref{fig:ground-2} and~\ref{fig:ground-folded} show the raw and phase-folded ground-based photometry obtained using SOAR and ULTRACAM, while Figs.~\ref{fig:ground-lomb} and \ref{fig:ground-lomb-zoom} show the Lomb-Scargle periodograms.
Observations were obtained in 2021 (soon after the \textit{TESS} observations) and in 2022 and 2023.
The three sets of observations highlight the changing profile of the variability in \starname.

The lightcurves from 2021 show a similar (though not identical) sawtooth-shaped profile as was observed from \textit{TESS}, though with a notably smaller amplitude ($\approx 1$ per cent, compared to the $\approx 4$ per cent seen in \textit{TESS}).
The high-frequency red noise known as `flickering' that is typical of accreting systems is also seen.
A periodogram of the data from 2021 shows a peak at the 23.07\,min period and its second harmonic, with no other significant peaks.

In 2022, the profile of the lightcurve appeared somewhat different.
The middle panel of Fig.~\ref{fig:ground-folded} shows the phase-folded data from three nights in March 2022.
As Fig.~\ref{fig:ground-lomb} shows, the strength of the second harmonic has significantly increased relative to the fundamental frequency of variation.
In some 2022 nights the fundamental frequency is not detected at all, which may be a result of the short observing windows of these observations (60--75\,min per night).

Most notably, the precisely measured \textit{TESS} period of 23.07172(6)\,min did not successfully phase-fold the 2022 ULTRACAM data across multiple nights such that their times of minimum light coincided.
Folding the data on the \textit{TESS} period induces a drift of 20 per cent of a phase cycle ($\approx 5$\,min) between the nights of March 05 and March 06.
This implies a significant change to the photometric period of at least $\approx 0.08$\,min between 2021 and 2022.
We found that a folding period of 22.07\,min was successful in aligning the lightcurve minimima across all three nights in 2022 March, but note that this is only one of a number of equally acceptable aliases.

In 2023, the profile of the lightcurve was different again.
As Fig.~\ref{fig:ground-folded} shows, the amplitude in 2023 March was $\approx 4$ per cent, similar to that of the \textit{TESS} lightcurve.
The folded lightcurve shows four peaks per phase cycle, resulting in a strong fourth harmonic of the dominant frequency (Fig.~\ref{fig:ground-lomb}).
In 2023, as in 2021 but not in 2022, phase-folding the data using the \textit{TESS} period of 23.07172(6)\,min successfully aligned the lightcurve minima across all three nights from 2023 March.

In Fig.~\ref{fig:ground-lomb-zoom} we show the combined Lomb-Scargle periodogram of data from two sets of consecutive nights, 2022 March 05--06 and 2023 March 08--10.
The strongest alias of the 2023 March data clearly agrees with the \textit{TESS} photometric period.
In 2022 March, the favoured values of the periodogram are shifted towards higher frequencies than the \textit{TESS} frequency, and no alias is aligned with the \textit{TESS} frequency.
We are therefore confident in claiming that the photometric period observed in 2022 March is different to the period observed in 2021 or 2023.
We also note that in 2023, the peaks of the periodograms in \filgs\ and \filus\ are aligned, while in 2022 they are offset by several aliases relative to each other; this may suggest that the apparent signal seen in 2022 is a combination of two signatures with different characteristic spectra.

Given the complex aliasing (seen in Fig.~\ref{fig:ground-lomb-zoom}) and the short photometric coverage, it is difficult to precisely quote the 2022 frequency of photometric variability.
We estimate by eye that the frequency lies somewhere in the range $70 \pm 8$ cycles day$^{-1}$, or $20.5 \pm 2.0$\,min, in the understanding that this is an overestimate of the uncertainty.

\subsection{Spectral Energy Distribution}
\label{sec:sed}

\begin{figure}
\includegraphics[width=\columnwidth]{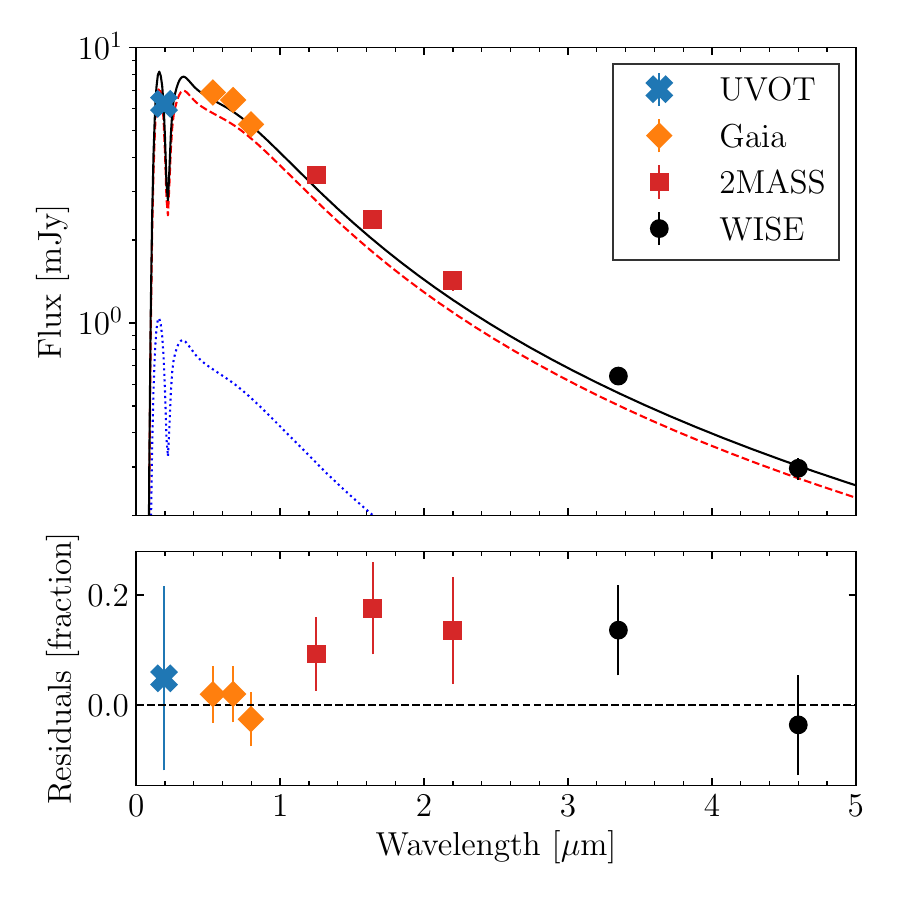}
\caption{Spectral energy distribution of \starname. The solid black line shows our best-fit, reddened SED model, which can be decomposed into its accretion disc component (red, dashed) and white dwarf component (blue, dotted).
The absorption-like feature at 2175\,\AA\ is an imprint of interstellar extinction.
}
\label{fig:sed}
\end{figure}

The Spectral Energy Distribution (SED) of \starname\ is shown in Fig.~\ref{fig:sed}.
In addition to the \textit{Swift}+UVOT data described in Section~\ref{sec:swift}, we retrieved data from the Galaxy Evolution Explorer \citep[\textit{GALEX};][]{Morrissey2007}, 
the American Association of Variable Star Observers (AAVSO) Photometric All-Sky Survey (APASS)\footnote{https://www.aavso.org/apass}, 
\textit{Gaia} Early Data Release 3 \citep{GaiaCollaboration2021}, 
the Two Micron All Sky Survey \citep[2MASS;][]{Skrutskie2006}, 
and the Wide-field Infrared Survey Explorer \citep[WISE;][]{Wright2010}.

There is a factor of two difference in flux between the \textit{Swift}+UVOT \textit{UVW2} measurement (observed 2021 July 27) and the archival GALEX NUV measurement (2011 May 11). 
The UVOT measurement is easier to reconcile with the optical flux measurements from \textit{Gaia}.
As the GALEX NUV measurement is based on only one observation, in which the target is somewhat close to the edge of the detector, we suggest that the measurement is affected by an unknown observational or calibration error.
Long-term variability of the source may also contribute to the difference between the UVOT and GALEX fluxes, but it is unlikely to be the entire explanation, because a factor of two change in flux would be unusual for a high-state AM\,CVn binary.
We note that there is also an offset between the optical flux measurements of APASS and \textit{Gaia}, but this smaller difference is easier to explain as resulting from variability of the source.
GALEX and APASS data were both excluded from the SED fitting process described below.

The SED cannot be described by a simple blackbody spectrum. 
This is expected for a high-state AM\,CVn binary, which is photometrically dominated by the accretion disc.

We modelled the SED of \starname\ using the method applied by \citet[][their Section 8]{Ramsay2018}.
It should be emphasized that this method utilizes a number of simplifying assumptions, and provides (at best) an approximation of the true SED.
The accretion disc was treated as a set of 200 linearly spaced, concentric, circular\footnote{This is a simplification, as in reality the presence of superhump variability implies that the disc is likely to be somewhat eccentric \citep{Patterson1993}.} annuli, extending from the surface of the accreting white dwarf to an outer disc radius $R_\mathrm{disc}$.
Each annulus was assumed to emit as a blackbody, with the temperature profile of the disc calculated in the standard manner for steady state accretion discs \citep[eg.][]{Warner1995}.
The central white dwarf was also treated as a blackbody emitter.
The white dwarf radius was calculated from its mass according to a typical carbon-oxygen core mass-radius relationship \citep{Verbunt1988} and its surface temperature was estimated from the mass transfer rate \citep[][their equation~1]{Bildsten2006}.
Free parameters in the model were the mass transfer rate \mdot, the primary mass $M_1$, the outer radius of the disc $R_\mathrm{disc}$, the orbital inclination $i$, the distance to the system $d$, and the interstellar reddening $E(B-V)$.

Gaussian priors were placed on $d$ \review{at $309.3 \pm 1.8$\,pc} (according to the \textit{Gaia} parallax measurement) and $E(B-V)$ \review{at $0.11 \pm 0.03$} \citep[according to the three-dimensional extinction maps of][]{Lallement2014,Lallement2018,Capitanio2017}\footnote{https://stilism.obspm.fr/}.
\review{We experimented with applying an additional Gaussian prior on $M_1$ \citep[$0.8 \pm 0.1 M_\odot$, following typical values for accreting white dwarfs, eg.][]{Pala2019}, and found that it did not significantly change the other derived parameters.}
Extinction was calculated using the extinction law of \citet{Fitzpatrick1999} with $R_V = 3.1$.
The outer disc radius was constrained to the range 0.07--0.11\,$R_\odot$, limits which were found by substituting reasonable ranges of primary and secondary masses into two common approximations for the outer disc radius: 30 per cent of the orbital separation and 70 per cent of the primary Roche lobe.

The model was converged on the data by minimising the $\chi^2$ using a Markov Chain Monte Carlo method \citep[MCMC;][]{Foreman-Mackey2013} with 32 walkers and 3000 iterations, which was sufficient for the chains to converge.
Best-fit values were determined from the median value of the last 500 iterations, and uncertainties determined as $1.4 \times$ their Median Average Deviation (MAD). 

It can be seen in Fig.~\ref{fig:sed} that the model is somewhat poorly fit to the data.
The best-fit model has $\chi^2 = 13.7$ with three degrees of freedom (reduced $\chi^2_\mathrm{red} = 4.6$).
The majority of the fits of \citet{Ramsay2018} have a somewhat better quality, but a minority of their fits are of a comparable quality\footnote{We make only a qualitative comparison as no goodness-of-fit parameter was quoted in that work.}.
The poorer fits, such as this one, are perhaps a result of the variability of the source between epochs of observation, or perhaps a result of a number of simplifying assumptions made during the modelling process (as detailed above).
\review{In particular, the excess in \textit{2MASS} and \textit{WISE} data may result from our overly-simple model of the accretion disc.}

The best-fit mass transfer rate was $\log (\dot{M} [\mathrm{M_\odot yr}^{-1}]) = -6.8 \pm 1.0$ \review{when the prior was applied to $M_1$, or $\log (\dot{M} [\mathrm{M_\odot yr}^{-1}]) = -7.0 \pm 0.8$ otherwise}.
When compared to other AM\,CVn binaries that were modelled in a similar way \citep{Ramsay2018}, the estimated mass transfer rate of \starname\ is the second-highest of all systems, beaten only by SDSS\,J1908+3940.
Note that, given the large uncertainties on our mass transfer rate estimate, it is only a 2\,$\sigma$ outlier from the general trend found by \citet{Ramsay2018}.
We also remark that AM\,CVn itself, perhaps the system most similar to \starname, was not modelled by \citet{Ramsay2018} due to the large scatter between its flux measurements. 

The best-fit value of $M_1$ \review{matched the applied prior. 
When run with no prior on $M_1$ except an upper limit of $1.4 M_\odot$, $M_1$ was essentially unconstrained.
The $E(B-V)$ found was $0.14 \pm 0.02$, also matching the applied prior.} 
The best-fit value of $i$ was $74 \pm 10 ^\circ$, although as with other results quoted in this section, it should be noted that the systematic uncertainties are not fully accounted for.

We experimented with adding a donor star into the SED model.
Donor stars are usually not visible in AM\,CVn binaries, though infrared excesses that may originate from the donor star have recently been observed in several systems \citep{Green2020,RiveraSandoval2021a}.
The donor star was approximated as a blackbody\footnote{This simplification was used because reliable spectral models of AM\,CVn donor stars do not currently exist.}, with its radius fixed to the secondary Roche lobe radius under an assumed donor mass of \review{0.125}\,$M_\odot$ \citep[the same donor mass as AM\,CVn itself;][]{Roelofs2006}, while the donor temperature was constrained to be less than 8000\,K.
The addition of a visible donor star slightly decreased the estimated mass transfer rate to $\log (\dot{M} [\mathrm{M_\odot yr}^{-1}]) = \review{7.0 \pm 0.8}$, and did not improve the quality of the fit sufficiently to justify the reduction in the degrees of freedom ($\chi^2 = 10.6$ and reduced $\chi^2_\mathrm{red} = 5.3$).
We therefore conclude that the donor star is not visible in \starname.

\subsection{Mass Transfer Rate from Bolometric Magnitude}

\review{
For comparison with the mass transfer rate derived in the previous section, we tested an alternate method proposed by \citet{Roelofs2007-HST}, in which the system luminosity is assumed to be completely dominated by accretion luminosity.
We note that this method systematically returns lower values of \mdot\ than the method of \citet{Ramsay2018}; \citet{Roelofs2007-HST} find values of \mdot\ for HP\,Lib and GP\,Com that are factors of $\approx 3$ and $\approx 10$ lower than those of \citet{Ramsay2018}.
}

\review{Following \citet{Roelofs2007-HST}, we used a bolometric correction of $-2.5 \pm 0.3$ (derived in that work under the assumption that the UV flux follows the spectral form of a 30\,000\,K blackbody) to find the bolometric magnitude from the $V$-band magnitude.
The mass transfer rate can then be derived by assuming that half of the difference in gravitational potential between the inner Lagrange point and the surface of the accretor is released as accretion luminosity.
We tested a number of trial values of $M_1$, $M_2$, and $\cos i$.
}

\review{
In this manner, mass transfer rates of $\log (\dot{M} [\mathrm{M_\odot yr}^{-1}])$ in the range $-8.4$ to $-9.4$ were found (best-fit value of $-8.9$), with the most critical unknown input being the value of $M_1$.
This value is similar to the values derived by \citet{Roelofs2007-HST} for the similar short-period system HP\,Lib. 
We note that the value derived for AM\,CVn itself in that work is larger, but this is likely due to the pre-\textit{Gaia} distance estimate being anomalously large \citep{Ramsay2018}.}

\review{
Due to the systematic uncertainties surrounding the bolometric correction, we favour the results of the SED fitting.
That being said, both methods have significant uncertainties and should be interpreted with caution.
}

\subsection{Long-Term Lightcurve}
\label{sec:longterm}

\begin{figure*}
\includegraphics[width=2\columnwidth]{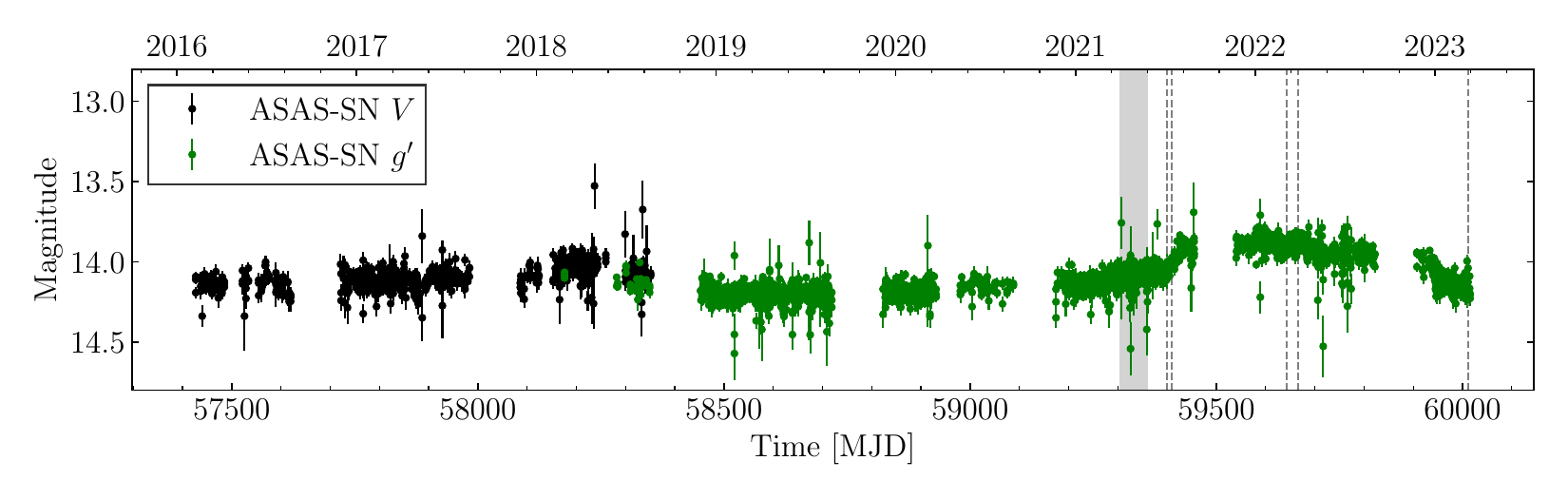}
\caption{Long-term lightcurve of \starname, showing an unusual long-term variation and the absence of outbursts.
The observation epoch of \textit{TESS} is shown by the shaded gray region, and those of ULTRACAM by dashed vertical lines.
}
\label{fig:longterm-lc}
\end{figure*}

In Fig.~\ref{fig:longterm-lc} we show photometry of \starname\ spanning 2000 days from February 2016 to March 2023, retrieved from the All-Sky Automated Search for SuperNovae \citep[ASAS-SN;][]{Shappee2014,Kochanek2017}.
The lightcurve shows an unusual long-term behaviour, in which the system grew brighter by $\approx 0.3$\,mag through 2021, remained bright for much of 2022, and faded again from late 2022 to early 2023.

Given that emission from the system is dominated by the accretion disc, the most likely explanation seems to be a temporary increase in the disc luminosity (due to either an increase in its temperature or radius, or some combination of the two).
Using the simple accretion disc model described in Section~\ref{sec:sed}, we estimate that a 0.3\,mag brightening in \filg-band magnitude would necessitate an increase in disc radius of $\approx 25$ per cent or an increase in $\log (\dot{M} / \mathrm{M_\odot yr}^{-1})$ of $\approx 0.4$\,dex.
For both of these estimates, all other parameters were held constant at their best-fit values.

Both the amplitude and timescale of the brightening event of \starname\ are somewhat similar to the `long outburst' phenomena observed in a number of long-period AM\,CVn binaries \citep{RiveraSandoval2020,RiveraSandoval2021a,Wong2021}.
Those phenomena have been suggested to result from a temporary increase in \mdot.
However, the nature of \starname\ as a high-state, disc-dominated AM\,CVn binary is quite different to the long-period, cold-disc AM\,CVn binaries in which such phenomena have been previously observed.
The `long outburst' phenomena are typically also associated with a reddening of the target, which is not seen here.
Comparing the \filus\filgs\filis\ ULTRACAM data, we do not find any measurable colour change between different epochs.
Note that this also suggests that a temperature change in the disc is unlikely.


\subsection{X-ray Detection}

X-ray emission in cataclysmic variables and AM\,CVn binaries originates from the boundary layer between the accretion disc and the accreting white dwarf \citep{Bath1974}, and possibly from a wind emitted by either the disc or the boundary layer \citep{Naylor1988}.
X-rays have been detected from a number of AM\,CVn binary systems across the entire range of orbital periods \citep{VanTeeseling1994,Ramsay2005,Ramsay2006,Ramsay2012a, Esposito2014,Wevers2016,RiveraSandoval2019,RiveraSandoval2020,RiveraSandoval2021a,Maccarone2023}.

\textit{Swift}+XRT detected 23 photons from \starname\ in 1240\,s of observation, leading to a count rate of $0.019 \pm 0.004$\,photons per second.
The spectrum was modelled using a power-law distribution.
\review{We used an absorption column density of $N_H = 0.96\times10^{21}$\,cm$^{-2}$, which is calculated from our fitted value $E(B-V) = 0.14$ (Section~\ref{sec:sed}) using a conversion factor $N_H = 2.21 \times 10^{21} A_V$\,cm$^{-2}$ \citep{Guver2009}, with $A_V = 3.1 E(B-V)$.
}

The resulting power law had a best-fit index \review{$\Gamma = 2.4 \pm 0.5$.} 
The observed model flux in the range 0.3--10\,keV is \review{$9.2 ^{+4.2}_{-1.8} \times 10^{-13}$\,erg\,cm$^{-2}$\,s$^{-1}$, equivalent to a luminosity of $1.1^{+0.5}_{-0.3}\times 10^{31}$\,erg\,s$^{-1}$ in the same energy range.}    
This value is in the typical range for AM\,CVn binaries, though with large uncertainties due to the low number of counts \citep{Ramsay2005,Ramsay2006,RiveraSandoval2019}.
We emphasize that the reliability of these results can be improved with further observations.


We performed a search of the 1240\,s of \textit{Swift}+UVOT photometry for variability, but given the short observing window which covered less than one orbital cycle, no meaningful upper limit could be derived.

\section{Spectroscopic Analysis}
\label{sec:spectroscopy}

\subsection{Average Spectrum}

\begin{figure*}
\includegraphics[width=2\columnwidth]{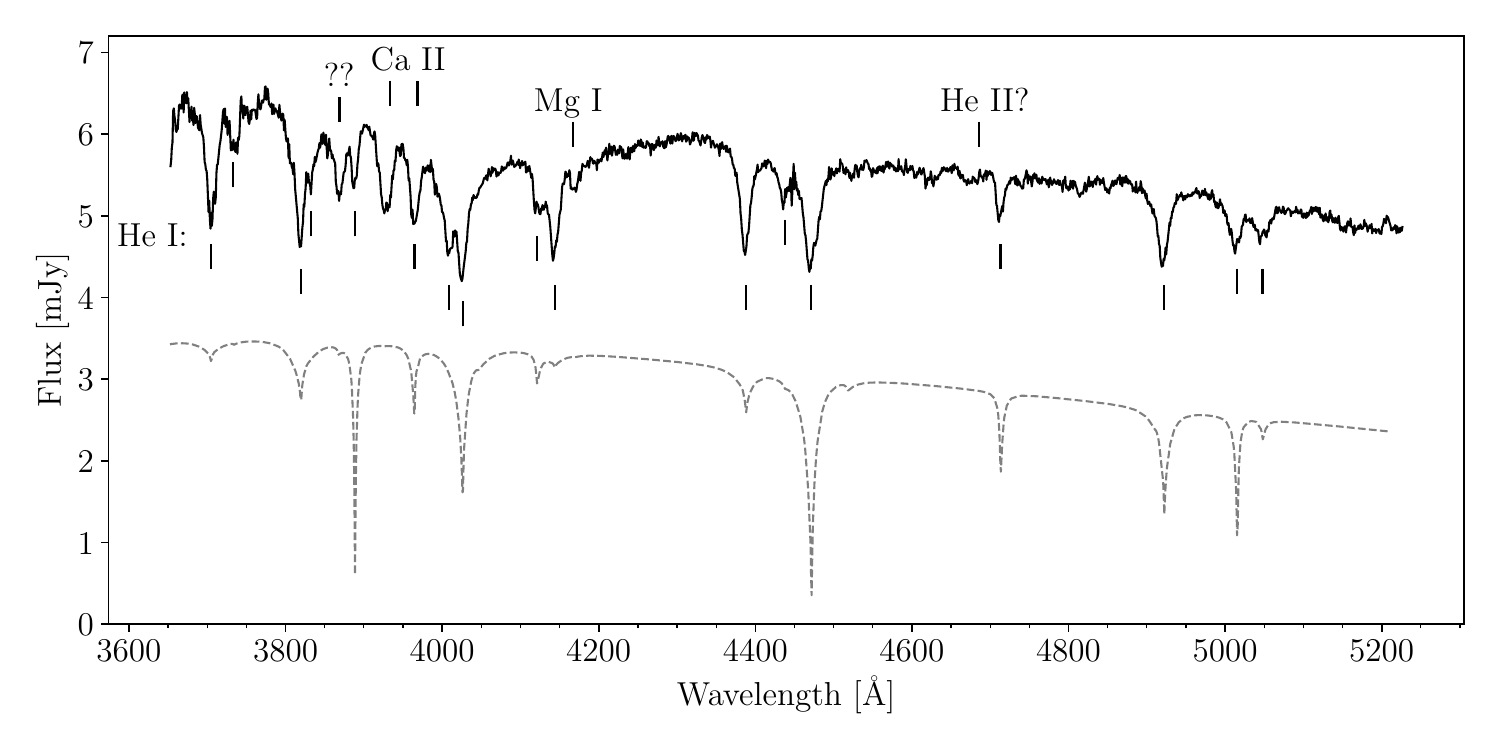}
\caption{Mean spectrum of \starname, combining 49 one-minute spectra obtained using Goodman+SOAR on 2021 July 05.
Visible features are marked.
All tick marks below the spectrum refer to \ion{He}{I}.
For comparison, we also plot a synthetic, 14\,000\,K, $\log g = 8.0$, helium-atmosphere white dwarf spectrum (grey, dashed) from  \citet{Cukanovaite2021} scaled to the same mean flux and offset by -3.5\,mJy.
}
\label{fig:av-spectrum}
\end{figure*}

\begin{figure}
\includegraphics[width=\columnwidth]{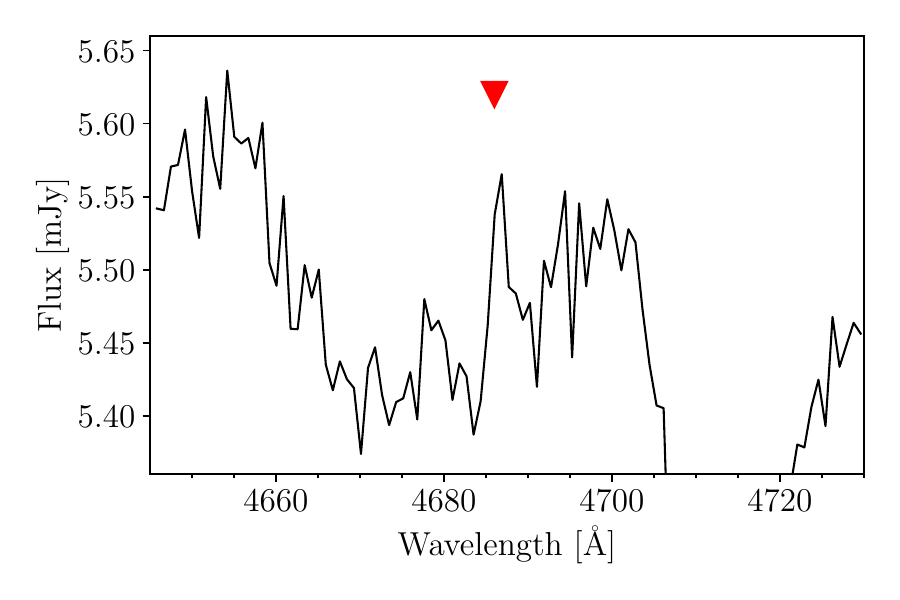}
\caption{Zoomed mean spectrum of \starname\ around the \ion{He}{II} 4686\,\AA\ line.
The expected wavelength of the line is shown with the red triangle. 
An emission line is marginally detected.
}
\label{fig:zoom-spectrum}
\end{figure}

\begin{table}
\caption{Equivalent widths of selected \ion{He}{I} lines in the mean spectrum of \starname.}
\begin{center}
\begin{tabular}{cc}
\label{tab:he-lines}
Wavelength [\AA] & EW [\AA] \\
\hline
3888.6 & $0.53 \pm 0.07$\\
4120.8 & $0.29 \pm 0.06$\\
4387.9 & $3.57 \pm 0.05$\\
4471.5 & $4.02 \pm 0.05$\\
4713.1 & $1.79 \pm 0.06$\\
4921.9 & $3.02 \pm 0.06$\\
5015.7 & $1.93 \pm 0.06$\\
\hline
\end{tabular}
\end{center}
\end{table}

The mean spectrum of \starname\ is shown in Fig.~\ref{fig:av-spectrum}.
It shows a blue continuum and a series of helium absorption lines, as is typical for a high-state AM\,CVn binary \citep[e.g.][]{Roelofs2006,Roelofs2007a,Fontaine2011,Kupfer2015}.
For comparison, we also show a synthetic, helium-atmosphere (DB) white dwarf spectrum with a temperature\footnote{The temperature was chosen by eye to be the best match to the observed spectrum, but this is not in any sense a representation of the true temperature of the AM\,CVn.} of 14\,000\,K and a surface gravity $\log g = 8.0$ \citep{Cukanovaite2021}. 
Although similar, the AM\,CVn spectrum has notable differences, in particular the ratios of relative depths of \ion{He}{I} lines.
The equivalent widths (EWs) of a selection of \ion{He}{I} line are listed in Table~\ref{tab:he-lines}.

When compared to the helium lines of AM\,CVn itself \citep{Roelofs2006}, the helium lines of \starname\ are somewhat narrower and significantly deeper. 
The helium line profiles of \starname\ each have a single minimum, while the line profiles of AM\,CVn itself each have two minima, resulting from the blue-shifted and red-shifted limbs of the accretion disc.
The narrow, single-core helium lines of \starname\ are more similar to those seen in another high-state AM\,CVn-type binary, SDSS\,J1908+3940 \citep{Kupfer2015}.
The single-core lines might suggest a relatively face-on inclination, although this would be at odds with the nearly edge-on inclination estimated from our SED fitting (Section~\ref{sec:sed}).

Emission from \ion{He}{II} is marginally detected (Fig.~\ref{fig:zoom-spectrum}). 
An absorption feature from \ion{Ca}{II} is seen at the K line (the \ion{Ca}{II} H line is not detectable due to blending with a nearby \ion{He}{I} line). 
\ion{Ca}{II} is also detected in AM\,CVn itself \citep[eg.][]{Patterson1993}.
A shallow absorption feature from 4660--4700\,\AA\ may originate from \ion{He}{II} or from a combination of \ion{He}{II} and \ion{Na}{I}, although the presence of \ion{Na}{I} would be surprising at the high temperatures implied by the presence of \ion{He}{II}.
There is an absorption line at approximately 3870\,\AA\ which we could not associate with any element typically seen in AM\,CVn binaries.

\subsection{Searching for Periodicity}

\begin{figure}
\includegraphics[width=\columnwidth]{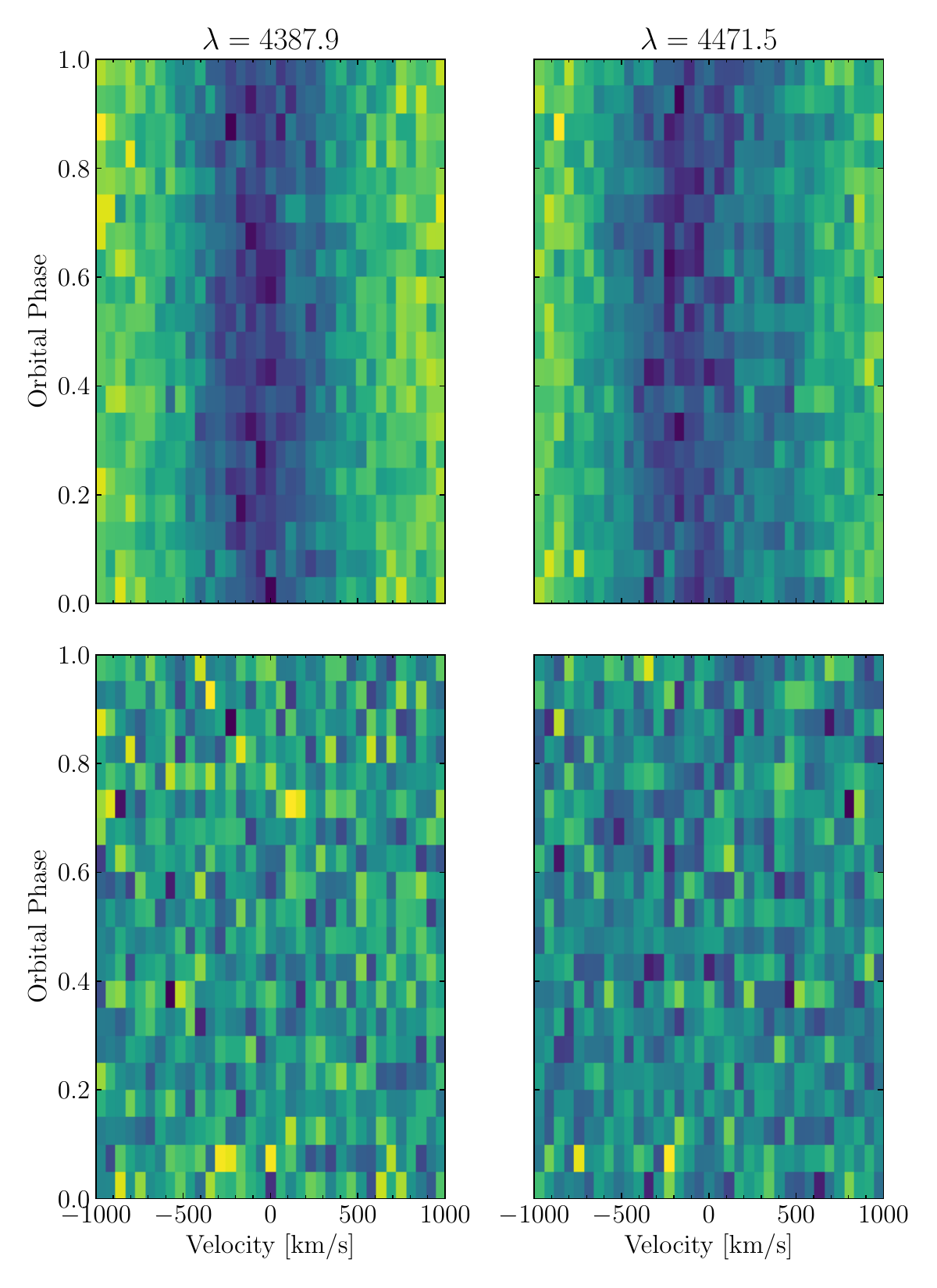}
\caption{\textit{Top row:} Trailed spectra of \starname\ from 2021 July 05, created using 49 back-to-back, one-minute exposures with Goodman+SOAR, phase-folded on the 23.07\,min period and showing the 4387.9\,\AA\ and 4471.5\,\AA\ \ion{He}{I} lines (left and right, respectively).
\textit{Bottom row:} The same trailed spectra with the mean spectrum (Fig.~\ref{fig:av-spectrum}) subtracted from each. We do not see any periodic feature in the residuals.
}
\label{fig:trails}
\end{figure}

In Fig.~\ref{fig:trails} we show trailed spectra of \starname, generated from the SOAR 1\,min spectra and phase-folded on the photometric period of 23.07 min.
No periodic pattern is seen in either the trailed spectra (upper panels) or the mean-subtracted trailed spectra (lower panels).
We repeated this process for every prominent \ion{He}{I} line and still found no evidence for any periodic spectral variability.

\review{As the orbital period is not necessarily the dominant photometric period, we also repeated this process for a further 200 frequencies, equally spaced between 60 and 80 cycles day$^{-1}$, so as to fully explore the range of possible orbital periods.
Folded trailed spectra were produced using each candidate period, for four spectral lines: \ion{He}{I} 4387.9\,\AA, \ion{He}{I} 4471.5\,\AA, \ion{Mg}{I} 4167.3\,\AA, and \ion{Ca}{I} 3933.7\,\AA.
No periodic signal was visible at any of the tested periods in any of the investigated lines.
}

Most AM\,CVn binaries show spectral emission from a bright spot feature, located at the intersection between the infalling accretion stream and the edge of the accretion disc, the radial velocity of which varies as a function of orbital phase \citep[e.g.][]{Kupfer2015}. 
The absence of such a feature in \starname\ is notable.
Such features are not usually difficult to detect in low-state systems, but can be much harder to detect in high-state systems, often requiring high-resolution observations \citep[e.g.][]{Roelofs2006}.
The system with the most similar spectrum, SDSS\,J1908+3940, shows periodic variability in only some of its absorption lines \citep{Kupfer2015}.

\begin{figure}
\includegraphics[width=\columnwidth]{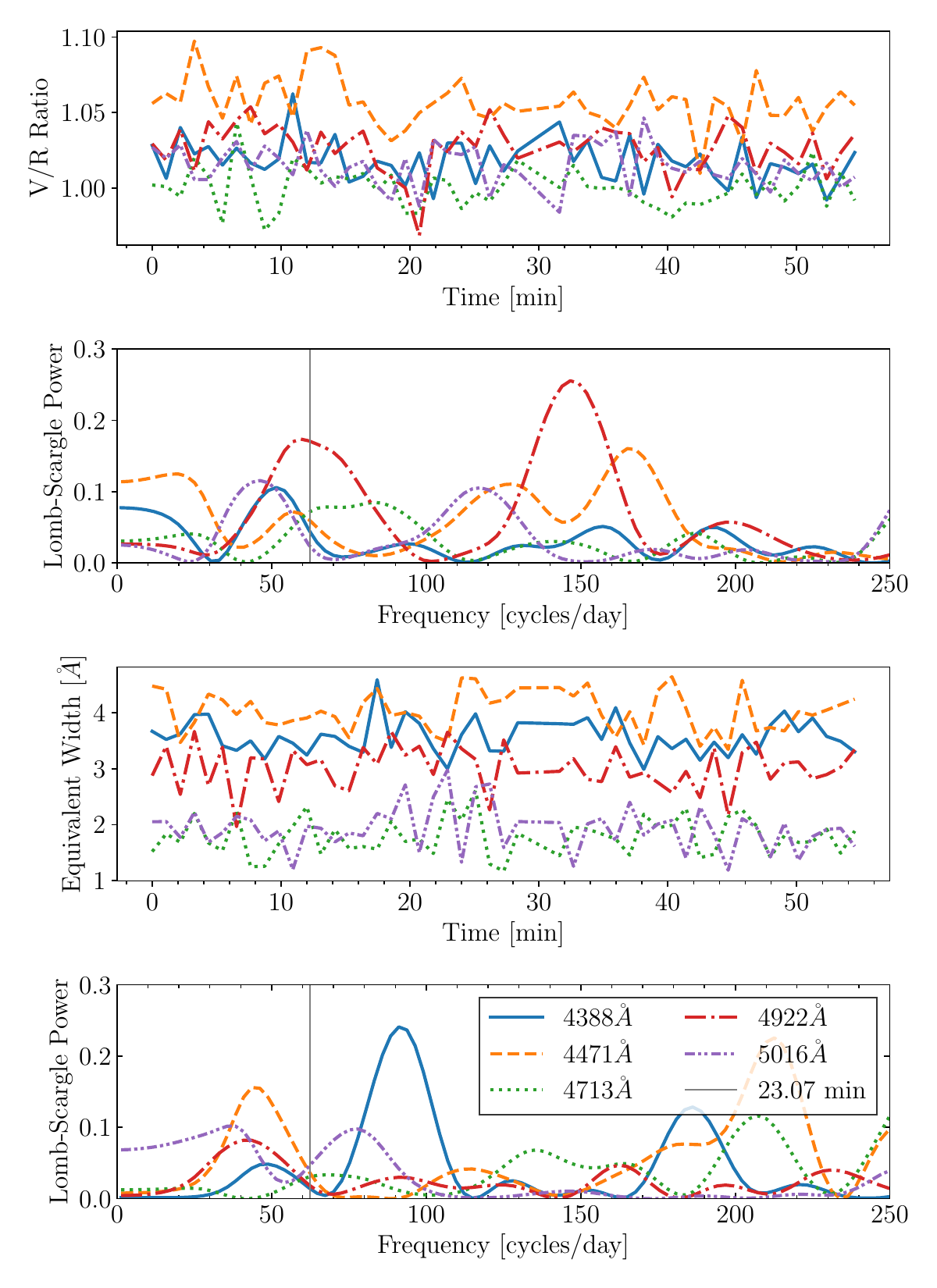}
\caption{Measured V/R ratios and equivalent widths of five \ion{He}{I} lines from the one-minute Goodman+SOAR spectra, and Lomb-Scargle periodograms of those measurements. On the periodograms, the 23.07\,min photometric period is marked with a dashed vertical line. In no line do we detect variability commensurate with the photometric periods.
}
\label{fig:vr-ew}
\end{figure}

\begin{figure}
\includegraphics[width=\columnwidth]{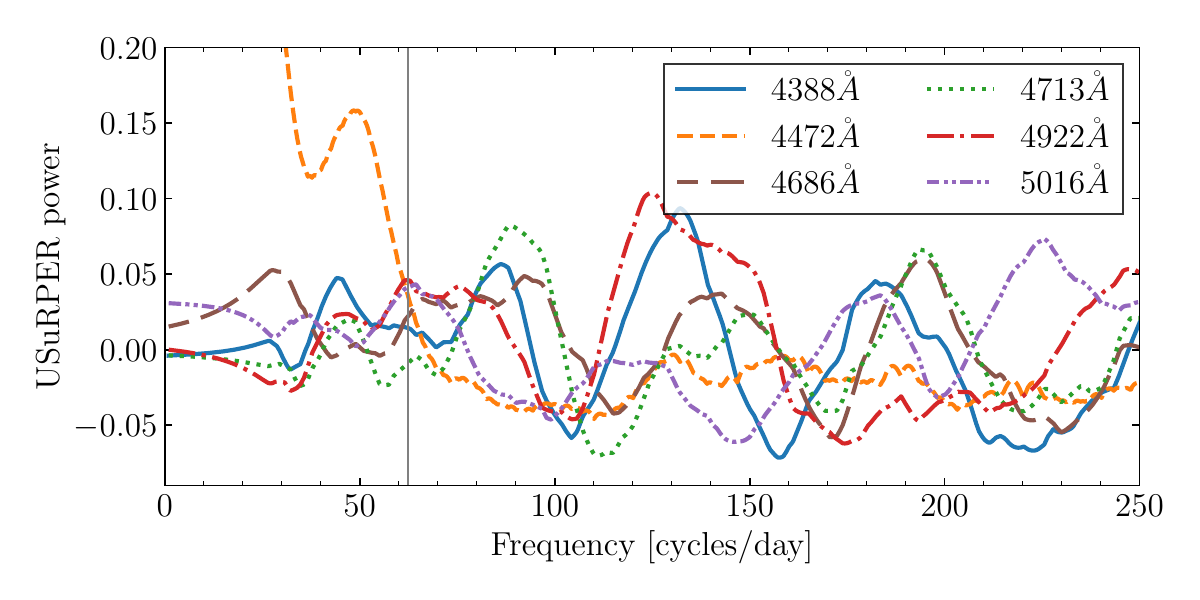}
\caption{Phase-distance correlation periodograms of five \hei\ lines and the \heii\ line at 4686\,\AA\ from the one-minute Goodman+SOAR spectra, calculated using the \software{usurper} algorithm.
}
\label{fig:usurper}
\end{figure}

We further searched for periodic patterns in two properties of the helium line profiles of \starname: the violet-to-red ratio (V/R) and equivalent width (EW) of each line.
The V/R ratio of a spectral line is the ratio of the integral under the violet wing of the line ($\lambda < \lambda_c$, where $\lambda_c$ is the central wavelength of the line as tabulated in Table~\ref{tab:he-lines}) to the integral under the red wing of the line ($\lambda > \lambda_c$).
For each of the 1\,min exposure time SOAR spectra, both V/R ratio and EW were measured from each of the five strongest helium absorption lines.
In Fig.~\ref{fig:vr-ew} we plot both properties over time, as well as a Lomb-Scargle periodogram of each. 
There is no periodicity that is significantly detected  (the most prominent peak, for the V/R ratio of the 4922\,\AA\ line, has a false alarm probability of nine per cent).

In addition, we searched for periodic variations in the shape of the spectral lines using the \software{usurper} algorithm (Unit-Sphere Representation PERiodogram), a form of phase-distance correlation periodogram \citep{Zucker2018,Binnenfeld2020}\footnote{Implemented using the \software{sparta} package (SPectroscopic vARiabiliTy Analysis) of \citet{Shahaf2020}.}.
The \software{usurper} algorithm searches for periodic changes in the overall shape of a set of input spectra, with no model-dependent assumptions about the underlying line shapes.
Once again we found no significant periodicity (Fig.~\ref{fig:usurper}).

\section{Discussion}
\label{sec:discussion}

\subsection{The Nature of the Photometric Periods}

In previously studied AM\,CVn binaries in high state or during outburst, the dominant photometric period is usually the superhump period \citep[e.g.][]{Roelofs2006,Armstrong2012,Isogai2019,Marcano2021}, although exceptions exist \citep{Kupfer2015}.
When the orbital and superhump period are both visible, it can be difficult to disentangle them without extended, continuous coverage \citep{Green2018b}.
As 23.07172(6)\,min is the only detected signal in the \textit{TESS} coverage of \starname, the most natural interpretation is that this is the superhump period.
Assuming a typical superhump excess of 1--3 per cent would imply an orbital period in the range of 22--23\,min.

The period appears constant through most of the \textit{TESS} data.
Towards the end of the coverage, a small-scale period change of order $\Delta P / P \sim 10^{-4}$ was detected.
Changes on a similar scale were detected in some of the periodic signals in SDSS\,J1908+3940 \citep{Kupfer2015},
\review{but not in the orbital signature. The higher harmonics of the signatures in \starname\ show changes in amplitude (Fig.~\ref{fig:tess-split-fold}), which were also seen in some periodic signals in SDSS\,J1908+3940, though \citet{Kupfer2015} do not comment on the harmonics of the orbital signature in particular.}

In 2022, \starname\ underwent two changes in its observed properties. 
Firstly, the system brightened by 0.3\,mag for the majority of 2022, before returning to its original brightness (Fig.~\ref{fig:longterm-lc}).
As discussed in Section~\ref{sec:longterm}, the change in brightness may be driven by a change in disc radius or temperature (the latter being closely linked to the mass transfer rate).
Secondly, the dominant photometric period changed from the period seen in \textit{TESS}, 23.07172(6)\,min, in July 2021 to a shorter period ($20.5 \pm 2.0$\,min) in March 2022, before returning to a period consistent with the \textit{TESS} period by the time of our March 2023 observations (Fig.~\ref{fig:ground-lomb-zoom}).
While the quoted uncertainties (limited by nightly aliases) are formally consistent between the two periods, the fact that no alias is consistent with the \textit{TESS} period suggests that there was at least some change in the period in 2022.
The closest alias to the \textit{TESS} period differs by 0.4 cycles day$^{-1}$ ($\Delta P / P \approx 0.006$).
The July 2021 observations were during the system's rise in brightness (Fig.~\ref{fig:longterm-lc}), while the March 2022 were during the brightness plateau, and March 2023 was following the descent back to its quiescent magnitude.
It seems natural to suggest there is a link between the two observed changes.

A change to the superhump period of this scale is not expected for a disc that is stably high or quiescent.
However, photometric period changes of a similar scale can be seen when a system is rising into outburst compared to the peak of the outburst \citep[`Stage A' compared to `Stage B' superhumps. eg.][]{Kato2009,Kato2013,Isogai2016}.
Those period changes are suggested to arise due to a change in the precession rate of the disc, which may for instance be driven by a change in which regions of the disc drive the overall precession rate \citep{Osaki2013}.
It may be that a similar change in disc state, related to the proposed increase in disc radius or mass transfer rate, might explain the difference in period between the 2022 dataset and the 2021 and 2023 datasets.

Alternatively, if the disc did indeed grow in radius during 2022, it may have entered a radius range in which the edge of the disc can be eclipsed by the donor star.
The phase-folded lightcurve from 2022 (Fig.~\ref{fig:ground-folded}) is somewhat similar in profile and amplitude to the disc-eclipsing binary PTF\,J1919+4815 \citep{Levitan2014}.
The inclination of PTF\,J1919+4815, 76--79$^\circ$, is consistent with the inclination of \starname\ that we estimate from our SED fitting ($74 \pm 10 ^\circ$, Section~\ref{sec:sed}).
Under this interpretation, the $20.5 \pm 2.0$\,min photometric period seen in 2022 would be the orbital period, or perhaps a blend of the unresolved superhump and orbital periods, as was seen by \citet[][]{Green2018b}.
The fact that periodograms of \filgs\ and \filis\ data from March 2022  peak at different frequencies (Fig.~\ref{fig:ground-lomb-zoom}), unlike those from March 2023 which are consistent between filters, also suggests a contribution from two unresolved periodic signals with different spectral profiles.

We consider the latter to be a more likely interpretation.
An orbital period in this range would be consistent with the 23.07\,min superhump period.
However, it is not possible to confirm this interpretation on the basis of the data that are currently available.
Further follow-up observations such as sustained, high-resolution spectroscopy, or photometry during a future brightening episode, may finally confirm the orbital period of this system.

\subsection{Gravitational Wave Radiation}

Given its (probable) short orbital period and close distance, the gravitational wave emission of \starname\ is likely to be strong.
Using the package \software{legwork} \citep{Wagg2022a,Wagg2022b}, we estimate the signal-noise ratio (SNR) that can be achieved for \starname\ using \textit{LISA}.
We adopted the estimated distance of $306.2 \pm 1.7$\,pc, orbital period of $20.5 \pm 1.5$\,min, orbital inclination of $74 \pm 10 ^\circ$, primary mass $0.8 \pm 0.1 M_\odot$, and secondary mass \review{$0.125 \pm 0.04 M_\odot$.}
The latter is \review{based on the secondary mass of AM\,CVn itself \citep{Roelofs2006}, with error bars large enough to include other typical values for} AM\,CVn binaries at this orbital period \citep[e.g.][]{Green2018b,vanRoestel2022}.

The expected SNR is \review{$2.8^{+1.5}_{-0.9}$, $4.9^{+2.4}_{-1.6}$, $24^{+15}_{-9}$, and $73^{+52}_{-25}$}, after 1, 2, 4, and 10 years of \textit{LISA} observations.
The dominant uncertainties come from the secondary mass, the orbital period, and the inclination.
Further observations may be able to more precisely measure these properties, and make \starname\ a valuable verification target for \textit{LISA} \citep{Kupfer2018,Kupfer2023}.

\section{Summary and Conclusions}
\label{sec:conclusions}

\starname\ is a bright system that shows photometric modulation on a period of 23.07\,min, first discovered in 2\,min-cadence \textit{TESS} observations. 
Its spectrum, photometric period, and accretion-driven photometric flickering support its classification as a high-state AM\,CVn binary system.
The 23.07\,min period is most likely to be the superhump period; if so, then the orbital period is likely to be in the range of 22--23\,min.
This makes \starname\ the shortest period binary system discovered by \textit{TESS} so far.
It is the third-brightest AM\,CVn binary system known, but has avoided detection until now, likely due to the absence of the photometric outbursts by which AM\,CVn binaries are often discovered.


During 2021--2022, the system underwent an unusual brightening event with an amplitude of 0.3\,mag which lasted for approximately one year.
At the same time, the dominant photometric period appears to have changed from 23.07\,min to a shorter period in the range 18.5--22.5\,min.
We propose an interpretation of these two changes in which an increase in the accretion disc radius (driven by an unknown cause) made the system brighter and caused the onset of disc-edge eclipses.
By early 2023, both the system brightness and the photometric period had returned to their previously recorded values.

Using an SED fit, we estimate that the mass transfer rate of the binary is unusually large ($\log (\dot{M} [\mathrm{M_\odot yr}^{-1}]) = -6.8 \pm 1.0$) when compared to other AM\,CVn binaries modelled in the same way.
This may help to explain how the system is able to sustain a high-state accretion disc at an implied orbital period of 22--23\,min.
A high mass-transfer rate may be driven by a donor star which is unusually high-mass or thermally inflated.

Given the nearby distance (306.2\,pc) and short orbital period (22-23\,min), \starname\ is likely to be a detectable source of gravitational waves with planned space-based detectors.
We estimate that \textit{LISA} is likely to be able to make a $3\sigma$ detection of \starname\ in its first two years of operation.
Further electromagnetic observations of \starname\ to fully characterise the system will make it a valuable verification binary system for \textit{LISA} \citep{Kupfer2018,Kupfer2023}.

It is remarkable that high-state AM\,CVn binary systems have now been serendipitously discovered in each of \textit{Kepler}, \textit{K2}, and \textit{TESS} \citep[][and this work]{Fontaine2011,Green2018b}.
Because high-state AM\,CVn systems do not undergo photometric outbursts, they are most easily found by short-cadence, high-precision photometric surveys.
The discovery of \starname\ reinforces the suggestion, implied by the distance distribution of AM\,CVn binaries \citep{Ramsay2018}, that a number of AM\,CVn binaries remain hidden even among bright and nearby stars.
Searches for blue, short-period variables in large photometric surveys \citep[see also][]{Burdge2020b,vanRoestel2022} are valuable tools to find these hidden AM\,CVn binaries.

\section*{Acknowledgements}

\review{We are grateful to the anonymous reviewer for feedback which has significantly improved the quality of the manuscript.}
MJG was supported by the European Research Council (ERC) under the European Union's FP7 Programme, Grant No. 833031 (PI: Dan Maoz). 
JJH acknowledges support through TESS Guest Investigator Program 80NSSC22K0737. 
This project has received funding from the European Research Council (ERC) under the European Union’s Horizon 2020 research and innovation programme (Grant agreement No. 101020057).
BNB acknowledges support through TESS Guest Investigator Programs 80NSSC19K1720 and 80NSSC21K0364.
TRM and IP were supported by the UK's Science and Technology Facilities Council (STFC), grant ST/T000406/1. IP additionally acknowledges a Warwick Astrophysics prize post-doctoral fellowship made possible thanks to a generous philanthropic donation.
VSD and ULTRACAM are funded by the UK’s Science and Technology Facilities Council (STFC), grant ST/V000853/1.
ASB was supported by the National Science Centre under project No.\,UMO-2017/26/E/ST9/00703. SGP acknowledges the support of a STFC Ernest Rutherford Fellowship. ADR and SOK were supported by Conselho Nacional de Desenvolvimento Cient\'{\i}fico e Tecnol\'ogico - Brasil (CNPq).
This research was supported in part by the National Science Foundation under Grant No.~NSF PHY-1748958 (Kavli Institute of Theoretical Physics).

This work made use of the software \software{iraf}, \software{xspec} and \software{daophot}, as well as the \software{python} packages \software{numpy, matplotlib, scipy, astropy, emcee} and \software{extinction}.

This paper includes data collected by the \textit{TESS} mission. Funding for the \textit{TESS} mission is provided by the NASA's Science Mission Directorate.
This research is based in part on observations obtained at the Southern Astrophysical Research (SOAR) telescope, which is a joint project of the Minist\'{e}rio da Ci\^{e}ncia, Tecnologia, e Inova\c{c}\~{a}o (MCTI) da Rep\'{u}blica Federativa do Brasil, the U.S. National Optical Astronomy Observatory (NOAO), the University of North Carolina at Chapel Hill (UNC), and Michigan State University (MSU). 
This work is based in part on observations collected at the European Organisation for Astronomical Research in the Southern Hemisphere under ESO programmes 105.D-0761, 0108.D-0718, and 0110.D-4325.
This research made use of the AAVSO Photometric All-Sky Survey (APASS), funded by the Robert Martin Ayers Sciences Fund and NSF AST-1412587.

\section*{Data Availability}

The 2\,min \textit{TESS} data used in this work are publicly available via the Barbara A. Mikulski Archive for Space Telescopes (MAST). 
SED data displayed in Fig.~\ref{fig:sed} are also publicly available.
Other data will be made available on reasonable request to the authors.




\bibliographystyle{mnras}
\bibliography{refs} 






\bsp	
\label{lastpage}
\end{document}